\documentclass{ecai} 

\usepackage{latexsym}
\usepackage{amssymb}
\usepackage{amsmath}
\usepackage{amsthm}
\usepackage{booktabs}
\usepackage{enumitem}
\usepackage{graphicx}
\usepackage{color}

\usepackage{mathtools}
\usepackage{bbm}
\usepackage{pifont}

\newtheorem{theorem}{Theorem}
\newtheorem{lemma}[theorem]{Lemma}

\newtheorem{definition}{Definition}

\newtheorem{example}{Example}

\usepackage{colortbl}
\usepackage{tikz}
\newcommand*\circled[1]{\tikz[baseline=(char.base)]{
            \node[shape=circle,draw,inner sep=0.1pt] (char) {#1};}}

\newtheorem{asu}{Assumption}
\newcounter{subassumption}[asu]

\definecolor{lightlime}{HTML}{DFFFD6} 
\definecolor{darklime}{HTML}{90FF90} 
\definecolor{darkerlime}{HTML}{50FF50} 
\usepackage[caption=false, font=footnotesize]{subfig}

\newcommand*{\Resize}[2]{\resizebox{#1}{!}{$#2$}}%
\newcommand{\fref}[1]{Figure~\ref{#1}}
\newcommand{\tref}[1]{Table~\ref{#1}}
\newcommand{\sref}[1]{Section~\ref{#1}}
\usepackage[misc]{ifsym}

\newcommand{\BibTeX}{B\kern-.05em{\sc i\kern-.025em b}\kern-.08em\TeX}

\newcommand\HUGE{\fontsize{17.3}{25}\selectfont}

\begin{document}

\begin{frontmatter}




\title{  {\large \textit{Appearing in the Proceedings of the 28th European Conference on Artificial Intelligence (ECAI 2025)}}\\[1ex]
\HUGE{ When Secure Aggregation Falls Short: Achieving Long-Term Privacy in Asynchronous Federated Learning for LEO Satellite Networks}}


\author[A] {\fnms{Mohamed Elmahallawy}\thanks{Corresponding Author. Email: mohamed.elmahallawy@wsu.edu.}}
\author[B]{\fnms{Tie Luo}} 

\address[A]{School of Engineering \& Applied Science, Washington State University, Richland, WA 99354, USA }

\address[B]{Department of Electrical and Computer Engineering, University of Kentucky, Lexington, KY 40506}


\begin{abstract}
Secure aggregation is a common technique in federated learning (FL) for protecting data privacy from both curious internal entities (clients or server) and external adversaries (eavesdroppers). However, in dynamic and resource-constrained environments such as low Earth orbit (LEO) satellite networks, traditional secure aggregation methods fall short in two aspects: (1) they assume continuous client availability while LEO satellite visibility is intermittent and irregular; (2) they consider privacy in each communication round but have overlooked the possible privacy leakage through multiple rounds. To address these limitations, we propose \textbf{LTP-FLEO}, an \textit{asynchronous FL framework} that preserves \textit{long-term privacy} (LTP) for LEO satellite networks. LTP-FLEO introduces (i) \textit{privacy-aware satellite partitioning}, which groups satellites based on their predictable visibility to the server and enforces joint participation; (ii) \textit{model age balancing}, which mitigates the adverse impact of stale model updates; and (iii) \textit{fair global aggregation}, which treats satellites of different visibility durations in an equitable manner. Theoretical analysis and empirical validation demonstrate that LTP-FLEO effectively safeguards both model and data privacy across multi-round training, promotes fairness in line with satellite contributions, accelerates global convergence, and achieves competitive model accuracy.

\end{abstract} 

\end{frontmatter}

\section{Introduction}\label{sec:intro}

Federated Learning (FL) has emerged as an innovative distributed machine learning (ML) paradigm  \cite{mcmahan2017communication} that can be exploited for the Internet of Remote Things (IoRT) \cite{IoRT2023olive,IoRT2022tibet}. FL empowers devices within the IoRT to collaboratively train large-scale ML models without needing their respective private data leaving each device. The FL framework typically operates as follows: (1) Each client (an IoRT device) independently trains a local ML model using its private data; (2) The client uploads the trained model parameters to an aggregation server ($\mathcal{AS}$); (3) The $\mathcal{AS}$ aggregates all the received client models into a global model, which is then redistributed to all the clients for retraining; (4) The above repeats iteratively for multiple rounds until the global model converges.
\begin{figure}[!t] 
     \centering
     \includegraphics[width=0.8 \linewidth]{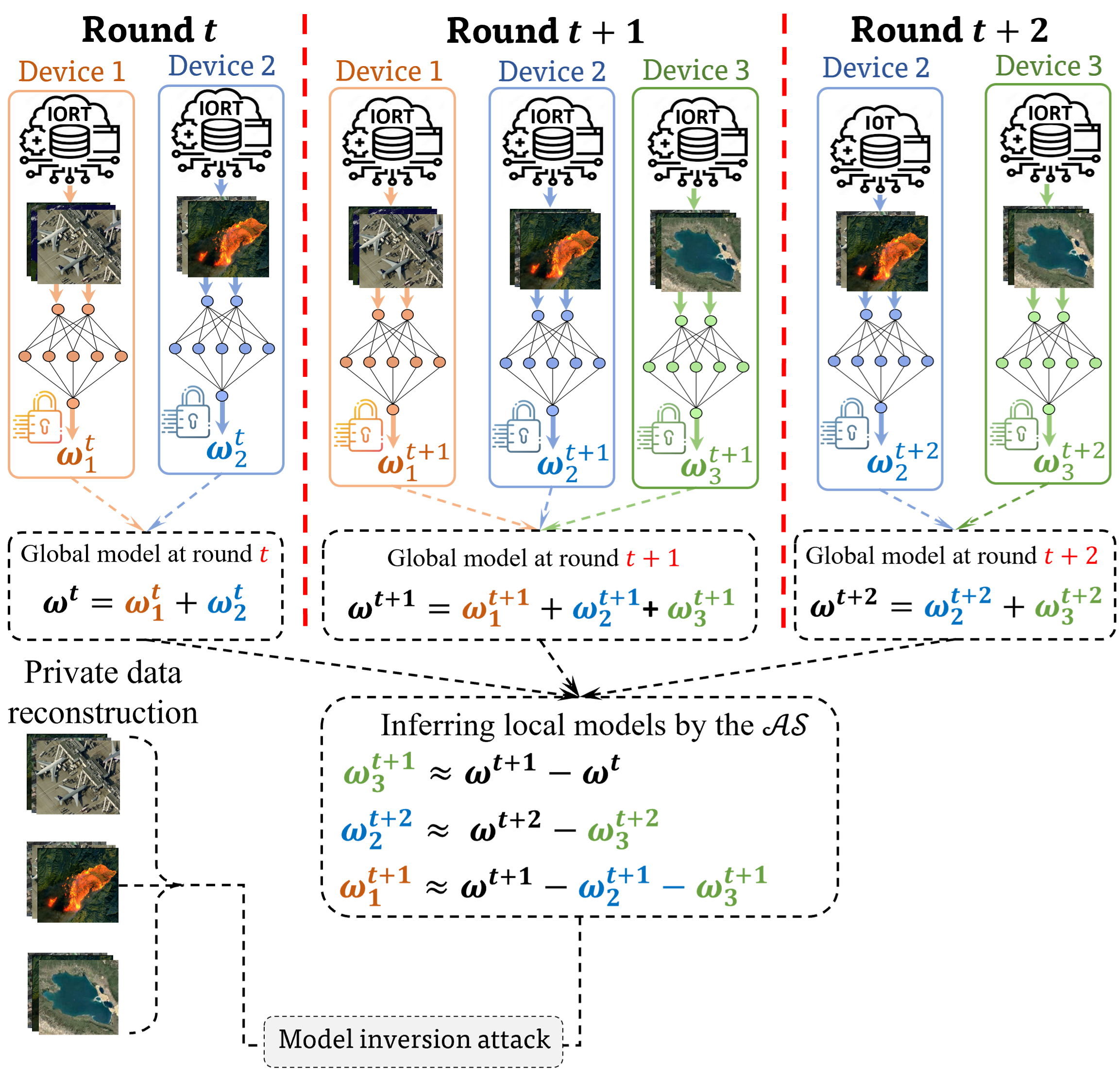}
    \caption{An illustration of the privacy leakage across multiple FL rounds. Note that this issue persists even if individual client models are encrypted.}
\label{fig:privace_breach}\vspace{-0.3cm}
\end{figure}

However, the communication channel between IoRT devices and the $\mathcal{AS}$ is usually insecure, exposing it to potential threats from curious insiders (such as $\mathcal{AS}$ or another device) or adversarial outsiders (eavesdroppers). These threats involve intercepting or stealing the model parameters transmitted over the channel, and reconstructing the original private data using techniques such as model inversion or membership inference attacks \cite{geiping2020inverting,nasr2019comprehensive}.


Secure aggregation techniques, such as homomorphic encryption (HE)~\cite{ma2022privacy}, multi-party computation (MPC)~\cite{kanagavelu2020two}, and differential privacy (DP)~\cite{wei2020federated}, have been adopted to safeguard client data privacy in FL. However, HE and MPC usually incur substantial computational and communication overhead, while DP can significantly degrade model performance due to its introduced noise. More importantly, these privacy-preserving mechanisms offer only \emph{short-term privacy} (STP), which secures data within a single communication round but not necessarily across multiple rounds. This problem is especially amplified in FL systems where clients are intermittently available~\cite{xiang2024efficient}, such as in low Earth orbit (LEO) satellite networks. Due to sporadic and irregular connectivity with the ground station (GS), LEO satellites are unable to participate consistently in each communication round, increasing their susceptibility to privacy leakage over time \cite{elmahallawy2023secure}. This constraint opens the door for strategic adversaries to exploit cross-round vulnerabilities and infer sensitive information, despite the use of traditional secure aggregation techniques.

To illustrate this privacy leakage over multiple FL rounds, we provide an example in \fref{fig:privace_breach}. At round $t$, the set of IoRT clients $\mathcal S^{t}=\{1,2\}$ are visible to the $\mathcal{AS}$ and hence participate by sending their encrypted models to $\mathcal{AS}$. The global model is therefore aggregated as $\boldsymbol{w}^{t}=\boldsymbol{w}^{t}_{1}+\boldsymbol{w}^{t}_{2}$ (skipping weighted sum for simplicity). In the next round $t+1$, supposing the available client set becomes $\mathcal S^{t+1}=\{1,2,3\}$, the global model will be aggregated as $\boldsymbol{w}^{t+1}=\boldsymbol{w}^{t+1}_{1}+\boldsymbol{w}^{t+1}_{2}+\boldsymbol{w}^{t+1}_{3}$. Typically, a local client model does not vary much between two consecutive rounds (i.e., $\boldsymbol{w}^{t}_{1}\approx\boldsymbol{w}^{t+1}_{1}$,  $\boldsymbol{w}^{t}_{2}\approx\boldsymbol{w}^{t+1}_{2}$, and $\boldsymbol{w}^{t}_{3}\approx\boldsymbol{w}^{t+1}_{3}$) especially in later training rounds. Hence, if the $\mathcal{AS}$ is curious about clients' private data, it can obtain an individual model (in this case client 3's) by $\boldsymbol{w}^{t}_{3}\approx\boldsymbol{w}^{t+1} - \boldsymbol{w}^{t}$, and subsequently reconstruct the private data using model inversion \cite{liu2023long,so2023securing}. Note that encrypting individual client models $\boldsymbol{w}_{1,2,3}$ using techniques such as HE or MPC~\cite{ma2022privacy,kanagavelu2020two}, or injecting noise using DP~\cite{wei2020federated}, does not solve the problem. This is because the aggregated global model is {\em plaintext} despite that individual local models are in {\em ciphertext}, which is made feasible by any  cryptography techniques.

This problem could arise in all FL scenarios with {\em dynamic} client participation, where only a subset of clients participate in each round, for example in asynchronous FL and cross-device synchronous FL,

This challenge highlights the need for privacy preservation \emph{across multiple communication rounds}, which we define as \textbf{long-term privacy} (LTP). LTP is a stronger concept than STP in that LTP subsumes STP and preserving the former automatically preserves the latter.

In this paper, we address the challenge of protecting LTP in LEO satellite networks---a representative and highly dynamic IoRT environment. We choose this setting because: (1) It has attracted significant industrial investment from companies such as SpaceX, OneWeb, and Amazon~\cite{rossi2025resource}, while its integration with FL represents an emerging technological trend that has sparked substantial research interest; (2) the risk of LTP leakage is more pronounced because the ``straggler'' problem is more severe in LEO environments---a varying set of satellites are hidden from the $\mathcal{AS}$ for a long time before becoming visible; and (3) the confidentiality of satellite data is particularly crucial as such data are often linked to homeland security ~\cite{routray2020military}.


To this end, we propose \textbf{LTP-FLEO} (\textbf{\underline{L}}ong-\textbf{\underline{T}}erm \textbf{\underline{P}}rivacy-preserving asynchronous \textbf{\underline{F}}ederated learning for \textbf{\underline{L}}ow \textbf{\underline{E}}arth \textbf{\underline{O}}rbit satellite networks), a framework designed to preserve privacy of individual satellite data across any number of FL communication rounds. It achieves this goal via three design components: (1) {\em Privacy-aware satellite partitioning}, to ensure the $\mathcal{AS}$ cannot infer any information about individual satellite models, either in a single or across multiple FL rounds; 
(2) {\em Model age balancing}, to reduce the adverse impact of stale client model updates on the global model; (3) {\em Fair global aggregation}, to ensure satellites in different orbits contribute equitably to the global model based on the quantity and quality of their data. The contributions of this work are:
\begin{itemize}[leftmargin=*,topsep=1pt]
    \item To the best of our knowledge, this is the first work to provide an LTP guarantee for asynchronous FL in LEO satellite networks. This setting is a natural and well-motivated choice, as LEO networks are characterized by highly irregular client availability and intermittent connectivity---conditions under which asynchronous FL is particularly well-suited. That said, the core principles behind LTP-FLEO are broadly applicable to asynchronous FL; the LEO environment simply serves to amplify their relevance and impact.
    
    \item We provide theoretical guarantees for LTP and conduct a convergence analysis of LTP-FLEO. Our analysis shows that the proposed mechanisms---privacy-aware satellite partitioning, model age balancing, and fair global aggregation---jointly achieve not only strong privacy protection but also convergence and fairness at the global level.
    \item Through extensive simulations on diverse datasets---including a real-world satellite dataset with non-IID distributions---we validate our theoretical findings. LTP-FLEO is shown to deliver strong LTP guarantees without introducing any additional computational or communication overhead. It also reaches high classification accuracy (95\%) within just 5 hours of satellite-to-$\mathcal {AS}$ communication including training time, and maintains class-level fairness.
\end{itemize}
\vspace{-3mm}

\section{Related Work}
\vspace{-1mm}
\subsection{Secure Aggregation in General FL}
Secure FL aggregation has been a central focus of research, aiming to mitigate privacy and security concerns inherent in collaborative ML training against various adversaries such as eavesdroppers or curious servers/clients. Various techniques have emerged to achieve secure aggregation, including HE \cite{bonawitz2017practical,ma2022privacy}, MPC \cite{kanagavelu2020two,dong2023meteor}, and DP \cite{wei2020federated,chen2022fundamental}. HE enables computation on encrypted data, facilitating privacy-preserving aggregation without disclosing individual contributions. MPC allows multiple parties to jointly compute a function while preserving their inputs' privacy. DP introduces randomness/noise to the computation process to safeguard individual data privacy.  However, these approaches usually suffer from high computation and communication overheads, or they struggle to balance between adding noise and achieving fast convergence. Most importantly, all of them fail to achieve LTP guarantees, leaving the clients' models and their associated private data at risk of disclosure after a few communication rounds.

Recent works have taken initial steps toward preserving the privacy of transmitted client models across multiple rounds to ensure LTP. For instance, in \cite{chang2023privacy,wu2023esafl}, the authors proposed innovative methods leveraging HE or MPC to execute the entire FL process on encrypted data and models until the convergence of the global model, thereby safeguarding LTP. However, the incurred computation and communication overheads exceed the limits of typical IoT/IoRT devices, making these approaches impractical for resource-constrained devices. Furthermore, these schemes, when applied in a decentralized manner as FL, necessitate the transmission of private client data to non-colluding servers, contradicting the FL principle of preserving client data privacy and communication efficiency.

\vspace{-3.5mm}
\subsection{Secure Aggregation in FL-LEO Networks}
\vspace{-0.5mm}

While the integration of FL into LEO satellite networks is still in its early stages, some studies (e.g., \cite{10438925,shi2024satellite}) have proposed FL approaches to expedite the convergence speed of FL-LEO. 
However, these approaches overlooked the privacy and security risks associated with transmitting satellites' models over insecure channels. Moreover, the enhancement of FL's convergence speed in these studies relies on utilizing the inter-satellite link (ISL) among satellites, which is not available in many recently launched LEO networks \cite{lin2023fedsn}.

Some recent works, such as \cite{hassan2023sfl,elmahallawy2023secure}, have begun to address the risk of satellite model theft and model inversion (inferring private satellite data) in FL-LEO by employing inner product functional encryption (IPFE) or HE. Although these approaches have succeeded in reducing communication and computation overheads, they do not ensure LTP preservation. LTP can only be ensured if {\em all} the LEO satellites (IoRT devices) participate in every communication round (i.e, synchronous FL), which is impractical due to dynamic client participation and the possibility of clients dropping out in some rounds. This is particularly relevant in the context of LEO satellites, as they suffer from sporadic and intermittent connectivity with the GS, and are often available for communications for only a few minutes \cite{shi2024satellite}. 

On the contrary, this paper takes an asynchronous FL approach for LEO satellites, and without relying on ISL. Moreover, it also addresses the strategic LTP challenge.


\vspace{-3mm}

\section{System Model and Threat Model of LTP-FLEO}
\vspace{-1mm}
\subsection{System Model} We consider an LEO satellite network $\mathcal K$  comprising $K$ satellites indexed by $k=1,...,K$, distributed on $P$ orbits, each containing a number of equal-spaced satellites with an altitude of $h_p$ above the Earth's surface. Each satellite $k$ collects a dataset $\mathcal{D}_k=\{x_{k,j},y_{k,j}\}_{j=1}^{|\mathcal{D}_k|}$, consisting of $|\mathcal{D}_k|$ samples for a certain ML task, such as wildfire detection or border monitoring. $x_{k,j}$ is the $j$-th high-resolution image captured by satellite $k$, and $y_{k,j}$ is the label of $x_{k,j}$ which can be determined by methods such as \cite{ostman2023decentralised}. All satellites within the network $\mathcal K$ are collaborating on training a global ML model with the objective of minimizing the following loss function:\vspace{-0.2cm}
\begin{equation}
    \min_{\boldsymbol{w}\in\mathbb{R}^d}\{F(\boldsymbol{w})\}, ~\text{with}~F(\boldsymbol{w})\triangleq\frac{1}{|\mathcal{D}|}\sum_{k\in\mathcal{K}}|\mathcal{D}_k|F_k(\boldsymbol{w})
\end{equation} 
where $\boldsymbol{w}$ is a vector representing the global model parameters, $F_k(\boldsymbol{w})\triangleq\frac{1}{|\mathcal{D}_k|}\sum_{\{x_{k,j},y_{k,j}\}\in \mathcal{D}_{k}} f_{k}(\boldsymbol{w};x_{k,j},y_{k,j})$ is the satellite $k$'s loss function, and $f_{k}(\boldsymbol{w};x_{k,j},y_{k,j})$ is the training loss of the model $\boldsymbol{w}$ on the data point $\{x_{k,j},y_{k,j}\}$.

{\bf LTP-FLEO's Training Process.} LTP-FLEO takes an asynchronous FL training approach where only currently visible satellites will participate in updating the global model, without waiting for all the satellites in the LEO network to become visible to the $\mathcal{AS}$. Specifically, the training process of LTP-FLEO comprises 4 steps in each communication round $t$:

In Step \circled{1}, the $\mathcal{AS}$ broadcasts the current global model $\boldsymbol{w}^{t}$ (where $t=1,2,\dots$, with $\boldsymbol{w}^{1}$ is randomly initialized) to its currently visible satellites. In Step \circled{2}, each visible satellite trains the received $\boldsymbol{w}^{t}$ using its private data with a local optimizer such as SGD, and obtain an updated local model $\boldsymbol{w}_{k}^{t}$, through local epochs $i=1, \dots, I$ as $\boldsymbol{w}_{k}^{t,i+1}=\boldsymbol{w}_{k}^{t,i}- \eta_{t} \nabla F_{k} (\boldsymbol{w}_{k}^{t,i})$, where $\eta_{t}$ is the learning rate. Then, in Step~\circled{3}, each visible satellite encrypts its local model $\boldsymbol{w}_{k}^{t}$ using a secure aggregation scheme—such as IPFE as discussed in~\cite{elmahallawy2023secure}—and uploads the resulting ciphertext $\boldsymbol{C}_{k}^{t}$ to the $\mathcal{AS}$ for aggregation. Finally, in Step~\circled{4}, the $\mathcal{AS}$ aggregates the received encrypted models. Typically, $\mathcal{AS}$ can perform functional aggregation (e.g., via functional encryption) without decrypting the individual local models. The global model in plaintext is then computed as: \(\boldsymbol{w}^{t+1}= \sum_{k\in \mathcal{K}'} \frac{|\mathcal{D}_k|}{|\mathcal{D}_{\mathcal{K}'}|} \boldsymbol{w}_{k}^{t},\) where $\mathcal{K}'$ denotes the set of currently visible satellites, and $|\mathcal{D}_{\mathcal{K}'}|$ represents the total data size of participating clients \footnote{Statistical meta-information such as $|\mathcal{D}_{k}|$ and $|\mathcal{D}_{\mathcal{K}'}|$ can be safely shared between the $\mathcal{AS}$ and the satellites without violating privacy~\cite{lipractical}.}. The $\mathcal{AS}$ then transmits the aggregated model $\boldsymbol{w}^{t+1}$ to the next set of visible satellites.

This cyclic process continues across rounds $T$ until the global model converges. Steps \circled{1} to \circled{4} resemble those of conventional asynchronous FL \cite{xie2020asynchronous}. However, LTP-FLEO redesigns Steps \circled{3} and \circled{4} with its components of privacy-aware satellite partitioning, model age balancing, and fair global aggregation (explained in \sref{Sec:Method}), to achieve LTP preservation while ensuring model convergence.
\vspace{-0.1cm}
\subsection{Threat Model}\label{Threat}

Following the previous research on securing the FL aggregation process (e.g., \cite{sosecure,elmahallawy2023secure}), we consider the GS (i.e., ${\mathcal {AS}}$) and LEO satellites as ``honest-but-curious'' participants. These entities adhere to the FL protocol but possess a level of curiosity aimed at learning sensitive information from each other's models. This threat also includes the potential for the ${\mathcal {AS}}$ to collude with a subset of satellites (e.g., those from a particular vendor), attempting to infer data owned by other vendors. Such inference can be performed using membership inference attack or model inversion attack techniques.

Prior secure aggregation schemes concentrate on protecting STP within each round $t$ for each satellite model $\boldsymbol{w}_{k}^{t}$. 
However, LTP across multiple rounds can be compromised and lead to potential reconstruction of private satellite data, as discussed in \sref{sec:intro}. In this paper, we focus on ensuring LTP protection (which subsumes short-term) of satellite models against curious ${\mathcal {AS}}$, satellites, or any adversarial eavesdroppers on the communication channel.
\vspace{-0.1cm}

\section{Proposed LTP-FLEO Framework}\label{Sec:Method}

LTP-FLEO is designed to ensure LTP for satellite models and data without compromising accuracy or impeding convergence. It also offers fair aggregation among all satellites' models. Before delving into the details of the three main components of the LTP-FLEO framework, we first provide a formal definition of LTP for clarity:

\begin{definition}
Let $\mathcal D=\{\mathcal{D}_{1},\mathcal{D}_{2},\dots,\mathcal{D}_{K}\}$ represent the data distributed among all the $K$ clients in an FL setting. Given a sequence of FL training rounds $t=1,2,\dots,T$, where $T$ is the total number of rounds, let $\boldsymbol{w}^t$ be the global model parameters at round $t$ (which are in plaintext), and $\boldsymbol{w}^t_{k}$ be the local model parameters of client $k$ at round $t$ (which are in ciphertext). We define LTP as a property that, after any number of training rounds, the information accessible to the $\mathcal {AS}$—namely, $\{\boldsymbol{w}^{t-1}_{k}\}_{k=1}^{K}$ used for secure aggregation and the resulting  $\boldsymbol{w}^t$—does not reveal any private information about any individual client model $\boldsymbol{w}^t_{k}$ or its corresponding dataset $\mathcal{D}_{k}$. Formally, we can express LTP as:
\begin{align*}
    &\forall~t\in \{1,2,\dots,T\}, \forall~k\in\{1,2,\dots,K\}:\\
&\Pr\left(\text{infer}(\boldsymbol{w}_k^t) \Big| \boldsymbol{w}^\tau|_{\tau=1}^{t-1} \right)\approx 0, \quad 
\Pr\left(\text{infer}(\mathcal{D}_k) \Big| \boldsymbol{w}^\tau|_{\tau=1}^{t-1}\right)\approx 0.
\end{align*}
This definition asserts that, at any round $t$, even if the $\mathcal {AS}$ has kept track of the entire history of global models $\boldsymbol{w}^\tau|_{\tau=1}^{t-1}$, the probability that the $\mathcal {AS}$ can infer the plaintext parameters of any client model $\boldsymbol{w}_k^t$ approaches zero. Likewise, the probability that $\mathcal{AS}$ can recover the raw data in any client dataset $\mathcal{D}_{k}$ is also approaching zero.
\end{definition}

\subsection{Privacy-Aware Satellite Partitioning}\label{sec:grouping}

Towards the overarching goal of ensuring LTP for each individual LEO satellite model and its associated data, this component clusters the satellites into equally sized partitions. It is important to note that the partition size plays a critical role: a larger partition size provides a stronger LTP guarantee, 
but also increases the waiting time for the $\mathcal{AS}$ to collect all models from a given partition, potentially slowing down the FL convergence. Therefore, our approach aims to strike the best trade-off between partition size and FL convergence speed while simultaneously ensuring robust LTP protection. Our simulation results, presented in \sref{Sec:evaluation}, further confirm that this clustering strategy maintains competitive model accuracy.


Without loss of generality, we partition all the LEO satellites in the network $\mathcal{K}$ based on {\em the prediction of their visibility to the $\mathcal{AS}$}. Specifically, our partitioning scheme works as follows: 
\begin{enumerate}[left=0pt]

\item \underline{\bf Initialization:} 
\begin{itemize}[left=0pt]
    \item Before starting the FL process, we leverage orbital mechanics to predict the visibility (contact time) of each satellite $k\in\mathcal{K}$ to the $\mathcal{AS}$ using the methodology outlined in \cite{ali1999predicting}.
    
    \item Then, we determine the overlap in visibility between satellites (identify satellites that are simultaneously visible to the $\mathcal{AS}$).

    \item Finally, based on the predicted contact time $p_{k}^{t}$ for each satellite $k \in \mathcal{K}$, we cluster the network $\mathcal{K}$ into $\frac{|\mathcal{K}|}{L}$ partitions, each of size $L$, denoted as ${\mathcal{G}} = \{G_1, G_2, \dots, G_{|\mathcal{K}|/L}\}$. Satellites within each partition are selected to have sufficient overlap in their visibility windows, ensuring coordinated participation. In each round, the satellites in any partition $G\in \mathcal{G}$ will either all participate in the FL training or abstain from participation altogether. This means that, even if the $\mathcal{AS}$ stores the entire history of global models across all training rounds, the most it can reconstruct is the aggregated sum of models within each partition $G$. Consequently, no information about any individual satellite's model is revealed, thereby preserving LTP. For instance, if $L = 2$, aggregation is only possible when both satellites in the partition successfully transmit their models to the $\mathcal{AS}$. In this case, the $\mathcal{AS}$ can reconstruct only the partial sum of the two satellite models, without gaining access to any individual model parameters. We refer to $L$ as the \emph{target LTP guarantee}, since larger values of $L$ correspond to stronger privacy preservation (cf. Figure~\ref{fig:EuroSat}), as the $\mathcal{AS}$ learns only aggregated results over larger subsets of satellites, making individual model inference increasingly difficult.


\end{itemize}
\item \underline{\bf During FL Training:}

\begin{itemize}[left=0pt]

\item In each FL round $t$, the $\mathcal {AS}$ selects one or more partitions of satellites to participate in training the global model. To choose these partitions, it first selects the partition whose last visible satellite becomes visible the earliest among all the partitions; in other words, this partition is
\begin{equation}\label{eq:early}
    {G}^* = \arg\min_{G\in\mathcal{G}} \, \max_{k\in {G}} p_{k}^{t}
\end{equation} 
where ${G}$ has a cardinality of $|\mathcal{G}|=L$ satellites. Then, the $\mathcal {AS}$ adds every other partition whose common visible window among its members overlaps with the common visible window of the first partition. These partitions are collectively denoted by ${\mathcal{G}'}$.


\item Once the $\mathcal{AS}$ receives all the models from these partitions in round $t$, 
it employs the components discussed in \sref{sec:Optimal_selec} and \sref{sec:fair_agg} to select and aggregate these models.

\item This process will iterate over all the rounds until the FL global model converges at round $T$.
\end{itemize}
\end{enumerate}
\begin{figure}[!t]
     \centering
     \includegraphics[width=0.8\linewidth]{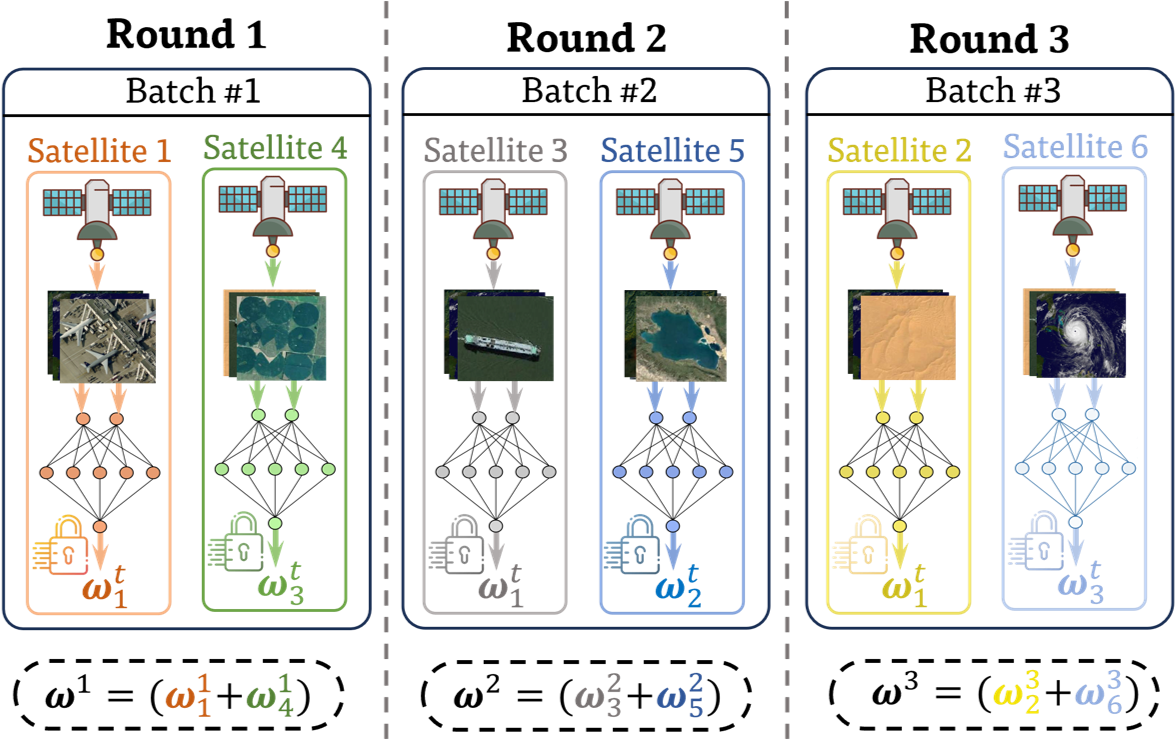}
    \caption{An illustration of LTP-FLEO's privacy-aware satellite partitioning component with $K=6$ and $L=2$.}
    \label{fig:group_scheme}
\end{figure}

{\example ($K=6, L=2$) In this example, we consider an LEO network consisting of 6 satellites as shown in \fref{fig:group_scheme}, which is partitioned into 3 partitions to ensure an LTP guarantee of $L=2$ (i.e., the $\mathcal{AS}$ can only obtain the partial sum of two models rather than any individual model in plaintext). In each FL round $t$, the $\mathcal{AS}$ has the flexibility to choose a set of partitions from $2^{\frac{|\mathcal{K}|}{L}}-1$ possible sets.  

The $\mathcal{AS}$ in this example selects only one partition in each round, assuming that only one partition of satellites is visible (or no other partitions overlap with this visible partition). 
However, even if more than one partition is visible and chosen, the target LTP guarantee of $L=2$ remains intact because the $\mathcal {AS}$ can only obtain a partial sum, at best, of two satellite models.} 
\vspace{-0.2cm}
\subsection{Model Age Balancing}\label{sec:Optimal_selec}

This component aims to optimize the partition selection during each round $t$ for participating in the global model aggregation. The main issue it addresses is to mitigate the impact of stale satellite models on the global model. Our insight is that straggler satellites with limited visibility to the $\mathcal{AS}$ could appear {\em early} in a {\em particular} round $t$ (hence selected by \eqref{eq:early}), but will appear less often across multiple rounds. Therefore, we let the $\mathcal {AS}$ utilize the historical participation frequency of each partition $G\in \mathcal{G}'$ to decide their participation in each round $t$. It prioritizes partitions with higher participation frequency to mitigate the impact of stale models from straggler satellites.

Recall the partitions ${\mathcal{G}'}$ selected for LTP preservation in \sref{sec:grouping}, which are the candidates for inclusion in the global model aggregation in round $t$. The $\mathcal {AS}$ computes the participation frequency $f_{{G}}^{t}$ for each partition ${G}\in \mathcal{G}'$ up to round $t-1$, as
\begin{equation}\label{eq:freuency}
    f_{{G}}^{t}\triangleq\sum_{m=1}^{t-1}\mathbbm{1} \Bigl\{s_{G}^{m}=1\Bigl\}, \, \forall {G}\in \mathcal{G}'
\end{equation}
where $s_{G}^{m}$ is a binary indicator that denotes whether the partition $G$ participated in round $m$ (recall that the member satellites of any partition either all participate or abstain altogether), and $\mathbbm{1}\{\cdot\}$ is the indicator function. Thus, given the candidate partitions $\mathcal{G}'$ and the historical participation frequencies $\boldsymbol{f}^{t}_{\mathcal{G}'} =(f_{{G}_1}^{t},f_{{G}_2}^{t},\dots,f_{G_{|\mathcal{G}'|}}^{t})$, the $\mathcal {AS}$ chooses only those partitions $G$ that meet the condition $t-\alpha \leq f_{{G}}^{t}\leq t-1$ to participate in global aggregation at round $t$. Here, $\alpha\geq 1$ denotes the tolerance factor used to control the acceptable level of staleness. The set of selected partitions is denoted by $\mathcal{G}''$. In the event that no partition satisfies the condition, the $\mathcal{AS}$ will skip this round. Below, we provide an example to illustrate how this component works.
{\example ($K=12$,~$L=3$, $t=10$, $\alpha=3$)  In this example, the 12 satellites are partitioned into 4 partitions according to their predicted visibility: ${G}_1=\{k_1,k_3,k_{8}\}$, ${G}_2=\{k_2,k_5,k_{10}\}$, ${G}_3=\{k_4,k_6,k_{11}\}$, and ${G}_4=\{k_7,k_9,k_{12}\}$. Suppose all of them are candidate partitions for aggregation at the round $t=10$, and the historical participation frequency vector is $\boldsymbol{f}^{10}=\big\{\underbrace{3}_{{G}_1},\underbrace{1}_{\tiny {{G}_2}},\underbrace{9}_{{G}_3}, \underbrace{7}_{{G}_4}\big\}$. 
According to the given tolerance value $\alpha$, the condition is $7 \leq f^{10}\leq 9$. Hence, the $\mathcal{AS}$ will select only ${G}_3$ and ${{G}}_4$ for participating in the global model aggregation at this round.}
 
\vspace{-0.2cm}
\subsection{Fair Global Aggregation}\label{sec:fair_agg}

This component aims to ensure fair aggregation among partitions. Although the selected partitions, as discussed in \sref{sec:Optimal_selec}, ensure the continual convergence of the global model across rounds (by favoring fresh models over stale models), they might introduce bias by favoring partitions with higher participation frequencies. 
To address this fairness issue, this component allows all partitions of satellites from different orbits to contribute to the global model in an equitable manner. 

Specifically, in each round $t$, once the $\mathcal {AS}$ has selected the partitions for aggregation (to update the global model $\boldsymbol{w}^{t+1}$), it assigns each partition a weight that takes into account the partition's participation frequency before round $t$ and the data size of the satellites within the partition, which is given by\vspace{-0.05cm} 
\begin{align}\label{eq:weighting}
\begin{split}
\beta_{{G}}^{t}=\frac{\gamma_{{G}}^{t}}{\sum_{{G}\in \mathcal{G}''}\gamma_{{G}}^{t}},\quad{\gamma_{{G}}^{t}}=\frac{f_{{G}}^{t}}{\sum_{{G}\in \mathcal{G}''} f_{{G}}^{t}}\times\frac{|D_G|}{|D_{\mathcal{G}''}|}
\end{split}
\end{align}
where $\beta_{{G}}^{t}$ is the weighting factor assigned to partition ${G}$, determined by considering the staleness of the partition and the total data size of satellites within the partition ${G}$, $\mathcal{G}''$ is the set of partitions selected by \sref{sec:Optimal_selec},  $|D_G|~\text{and}~|D_{\mathcal{G}''}|$ denote the total size of the satellite data in partition ${G}$ and in the set of selected partitions $\mathcal{G}''$, respectively.

Given \eqref{eq:weighting}, we can now relax the tolerance factor $\alpha$ to achieve a more balanced aggregation while mitigating model staleness at the same. For example, in an extreme case, we can set $\alpha=t$ so that even the lowest-participating partitions will be considered during aggregation, but contribute less by giving them a lower weight. In the meantime, the data size is also taken into account.

\begin{theorem}(Worst-Case Fairness Guarantee): 
In the worst-case scenario where the staleness tolerance factor, $\alpha=t$, allows the selection of all partitions while maintaining the target LTP guarantee, the weighting factor $\beta_{{G}}^{t}$ guarantees fair participation among all partitions, including those with lower participation rates, and also ensures the convergence of the FL global model. 
\begin{proof}
Assume all partitions ${G}\in\mathcal{G}'$ can be selected for aggregation at round $t$. In this scenario with $\alpha=t$, $\mathcal{AS}$ will select all partitions, including those with ``zero'' participation frequency. Our objective is to minimize the aggregation fairness gap, ensuring that all partitions contribute to the global model while mitigating the impact of staleness caused by partitions with lower participation on the convergence speed of the global model. This fairness gap $\mathsf{F}$ can be mathematically represented for any $\alpha$ as\vspace{-0.1cm}
\begin{align}
\mathsf{F}=\max_{{G}\in\mathcal{G}'} \limsup_{T\rightarrow\infty} \frac{1}{T}\mathbb{E}\biggr[\sum_{t=1}^{T}\mathbbm{1} \Bigl\{s_{G}^{t}=1\Bigl\}\biggr]-\\
    \min_{{G} \in\mathcal{G}'} \liminf_{T\rightarrow \infty} \frac{1}{T}\mathbb{E}\biggr[\sum_{t=1}^{T}\mathbbm{1} \Bigl\{s_{G}^{t}=1\Bigl\}\biggr]\nonumber
\end{align}
In this worst-case assumption with $\alpha=t$, $\mathsf{F}$ tends to zero as all partitions, including the partitions with no visible satellites at that round (where $\mathcal{AS}$ utilizes their last updated local model, which might be outdated), will be selected for aggregation. By utilizing \eqref{eq:weighting}, each partition is assigned a weight based on its staleness and the satellites' data size within it. This guarantees equitable participation across all partitions while reducing the staleness impact on the convergence of the global model.
\end{proof}
\end{theorem} \vspace{-0.1cm}
Once the $\mathcal{AS}$ computes $\beta_{{G}}^{t}$ for each partition, it aggregates them using the following equation
\begin{equation}\label{eq:aggre}
    \boldsymbol{w}^{t+1}=\sum_{{G}\in \mathcal{G}''} \beta_{{G}}^{t} \sum_{k\in {G}}\boldsymbol{w}_{k}^{t}
\end{equation}


\section{Theoretical Analysis}\label{Sec:theorical}
This section presents the theoretical analysis of LTP-FLEO regarding LTP guarantee and FL convergence.

\subsection{Long-term Privacy Guarantee of LTP-FLEO}

\begin{theorem}
Given  satellite network $\mathcal{K}$ and partition size $L$, LTP-FLEO ensures LTP with a level of guarantee $L$, while minimizing the aggregation fairness gap $\mathsf{F}$ for a given staleness tolerance factor $\alpha$.\vspace{-0.15cm}
\begin{proof}
According to our partitioning scheme described in \sref{sec:grouping}, satellites in the same partition either all participate in the global model aggregation or do not participate at all. Consequently, the $\mathcal{AS}$ is unable to reconstruct any individual model within a partition but only the partial sum of an entire partition. This can be seen by referring back to Eqn.~\eqref{eq:aggre}, where the aggregation implies that
\begin{align}
    &\boldsymbol{w}^{t+1}=\sum_{{G}\in \mathcal{G}''} \beta_{{G}}^{t} \sum_{k\in {G}}\boldsymbol{w}_{k}^{t} \\
    &=\beta_{{G}_1}^{t}\underbrace{\sum_{k\in {G}_1}\boldsymbol{w}_{k}^{t}}_{\text{Partition}_1}+\beta_{{G}_2}^{t}\underbrace{\sum_{k\in {G}_2}\boldsymbol{w}_{k}^{t}}_{\text{Partition}_2}+\dots+\beta_{G_{|\mathcal{G}''|}}^{t}\underbrace{\sum_{k\in {G_{|\mathcal{G}''|}}}\boldsymbol{w}_{k}^{t}}_{\text{Partition}_{|\mathcal{G}''|}}\nonumber.
\end{align}
With the condition that ${G}_a\cap {G}_b=\phi$ for $a\neq b$, it ensures that the $\mathcal{AS}$ can only reconstruct a partial sum of $L$ local models. 
Furthermore, $\beta_{G}^t$ leads to a fair participation ($\mathsf{F}$ is minimized when $\alpha=t$ where all partitions are eligible to participate with different weights) and mitigates the staleness impact on the FL global model by balancing participation frequency, model age, and data size.
    \end{proof}
\end{theorem}

\subsection{Convergence Analysis of LTP-FLEO}
For the convergence analysis of LTP-FLEO, we apply the following assumptions, commonly encountered in the literature \cite{Li2020On,ribero2022federated}, on the loss functions $F_1, \ldots, F_K$ for all satellites in our network $\mathcal {K}$.

\begin{definition}($\mathcal{L}$-Smooth)
    If a function $F$ satisfies $\|\nabla  F(\boldsymbol{a})-\nabla F(\boldsymbol{b})\|\leq \mathcal{L}\|\boldsymbol{a}-\boldsymbol{b}\|$, for any $\boldsymbol{a}$ and $\boldsymbol{b} \in \mathbb{R}^d$, we say that $F$ is $\mathcal{L}$-Smooth, where $\mathcal{L}$ is the Lipschitz constant.
\end{definition}
\begin{asu}[Smoothness]\label{as:1}
All the satellites' loss functions $F_1, \ldots, F_K$ in our network $\mathcal{K}$ are smooth such that $F_k(\boldsymbol{a})\leq F_k(\boldsymbol{b})+{(\boldsymbol{a}-\boldsymbol{b})}^{\top}\nabla F_{k}(\boldsymbol{b})+\frac{\mathcal{L}}{2}\|\boldsymbol{a}-\boldsymbol{b}\|^2_2$
\end{asu}
\begin{asu}[$\mu$-Strong Convex]\label{as:2}
All the satellites' loss functions $F_1, \ldots, F_K$ in our network $\mathcal{K}$ are strongly convex such that
$F_k(\boldsymbol{a})\geq F_k(\boldsymbol{b})+{(\boldsymbol{a}-\boldsymbol{b})}^{\top}\nabla F_{k}(\boldsymbol{b})+\frac{\mu}{2}\|\boldsymbol{a}-\boldsymbol{b}\|^2_2$
\end{asu}
\begin{asu}[Bounded variance]\label{as:3}
Let $\zeta_{k}^{t}={\{x_{j},y_{j}\}_{k}^{t}}$ denotes a data point randomly selected from the dataset $D_k$ of satellite $k$. This sampling is bounded by the variance of the stochastic gradients, applicable to all satellites in our network $\mathcal{K}$, such that $\mathbb{E}\|\nabla F_k(\boldsymbol{w}_{k}^t,\zeta_{k}^{t})-\nabla F_{k}(\boldsymbol{w}_{k}^t)\|_2^2\leq \sigma_{k}^{2}$
\end{asu}
\begin{asu}[Bounded stochastic gradients]\label{as:4}
For all satellites in our network $\mathcal{K}$, the expected square of stochastic gradients is uniformly bounded by $H$, satisfying the inequality condition $\mathbb{E}\|\nabla F_k(\boldsymbol{w}_{k}^t,\zeta_{k}^{t})\|_2^2\leq H^{2}$ 
\end{asu}

\begin{theorem}\label{th:Theorem}
Suppose Assumptions 1 through 4 hold, and the parameters $\mathcal{L}, \mu, \sigma_{k},$ and $H$ are defined. Consider an LEO network $\mathcal{K}$ partitioned into $|\mathcal{G}|$ partitions, with all satellites collaborating asynchronously to train an FL global model. LTP-FLEO selects a set of partitions $|\mathcal{G}''|$ in each round based on their participation frequency as in \eqref{eq:freuency} and weighted them to ensure fair participation as in \eqref{eq:weighting} for updating the global model using Equation \eqref{eq:aggre}, aiming for the convergence of the global model over $T$ rounds. Then, we have the following inequality: 
\begin{equation}
    \mathbb{E}[F(\boldsymbol{w}^{T})]-F^*\leq \frac{2\kappa}{\upsilon+IT}\Bigg(\frac{\lambda+\nu}{\mu}+\frac{\mu\upsilon}{4}\mathbb{E}\|\boldsymbol{w}^{1}-\boldsymbol{w}^{*}\|^2 \Bigg)
\end{equation}
where we set $\kappa=\frac{\mathcal{L}}{\mu}$, $\upsilon=\max\{8\kappa,I\}$, $\lambda=\sum_{k\in \mathcal{K}} {{s_k}}^2{\sigma_{k}}^2+6\mathcal{L} \Gamma+8(I-1)^2H^2$, $\nu=\frac{4I^2H^2}{K}$, and $\Gamma=F^*-\sum_{k\in \mathcal{K}}s_kF^{*}_{k}\geq0$, with $s_k$ and $F^*$ represent the satellite $k$ participation and the optimal value of $F$, respectively.

For determining the required number of rounds $T$ for the global FL model to achieve the target accuracy $\rho$, we have
\begin{equation}
\Resize{7.83cm}{T=\mathcal{O}\Bigg(\frac{1}{\rho}\biggl(\Big(1+\frac{1}{|\mathcal{G}|}\Big)IH^2+\frac{\sum_{k\in\mathcal{G}}s_k^2\sigma_k^2+\mathcal{L}\Gamma+\kappa H^2}{I}+H^2\biggl)\Bigg)}
\end{equation}
\end{theorem}

The proof of Theorem~\ref{th:Theorem} is provided in Appendix.

\begin{figure}[!t] 
     \centering
     \includegraphics[width=0.75\linewidth]{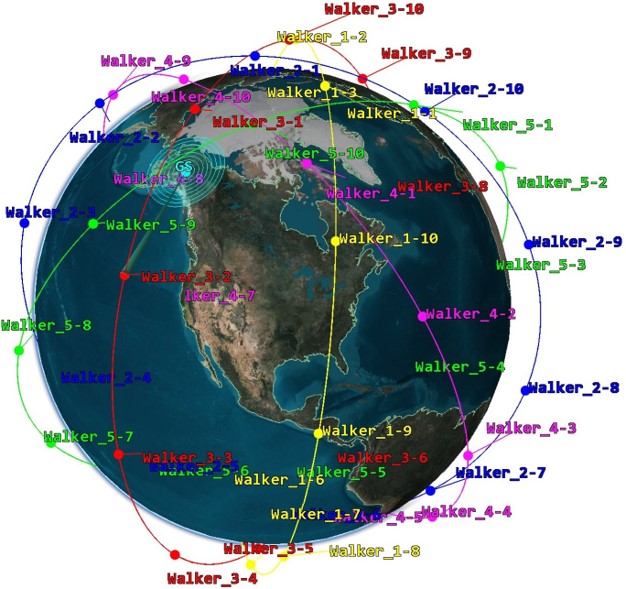}
    \caption{The Walker-Delta Satellite Network in our experiment.}
    \label{fig:walker}
\end{figure}
\begin{table}[!t]
\centering
\caption{Datasets, Models, and Training Parameters} 
\label{tab:1}
\resizebox{\linewidth}{!}{%
 \begin{tabular}{p{1.35cm}|p{0.9cm}|p{0.9cm}|p{1.23cm} |p{1.45cm}| p{0.95cm}|p{1.53cm}}
 \bottomrule 
Dataset &Size&Classes&Model&Parameters& Epochs& mini-batch\\
 \hline 
 MNIST &70,000&10&CNN&437,840&50&32\\
 \hline 
{\footnotesize CIFAR-10}&60,000&10&CNN&798,653&100&32\\
 \hline
 EuroSat &27,000&10&VGG-16&8,413,194&200&8\\
 \hline\hline
 \multicolumn{3}{c||}{$\eta=0.001-0.1$}&\multicolumn{4}{c}{SGD Momentum$=0.9$}\\
 \toprule
\end{tabular}}\vspace{-0.2cm}
\end{table}
\section{Experimental Results}\label{Sec:evaluation}
{\bf Experiment Setup.} We design a generic Walker-Delta satellite network (see \fref{fig:walker}) with an inclination angle of 80 degrees. This network comprises 5 orbits, each positioned at an altitude of 780 km above the Earth's surface (indicated by different colors in \fref{fig:walker}), and accommodates 10 equally spaced satellites. For our analysis, we specify the $\mathcal{AS}$ to be a GS located within a city in the USA, although it can be positioned anywhere on Earth, with a minimum elevation angle of 15 degrees.

To predict the visibility windows of satellites with the GS, we utilize the satellite simulation software Ansys STK \cite{ansys_stk}. For the remaining communication links, we adopt the configurations as detailed in \cite{elmahallawy2023secure}. In our training phase, we utilize the EuroSat dataset \cite{helber2019eurosat}, which comprises real satellite images captured by the European Space Agency's Sentinel-2 satellite missions and categorized into 10 different land covers, designed for classification tasks. We summarized all datasets and parameters used in the training phase in Table~\ref{tab:1}. Our model training was implemented using Python with PyTorch framework and run on the NVIDIA GeForce RTX 4090 GPU.

\begin{figure}[!t]
\centering
    {\subfloat[\fontsize{7.4pt}{9pt}\selectfont Ground Truth samples of EuroSat dataset.]
    {\includegraphics[width=0.4\textwidth]{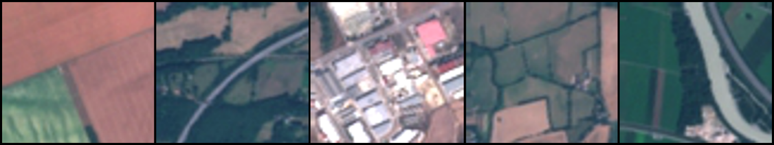}}}
    {\subfloat[\fontsize{7.4pt}{9pt}\selectfont Reconstructed samples of EuroSat dataset under FedSecure.]
    {\includegraphics[width=0.4\textwidth]{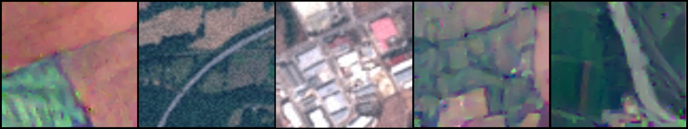}}}
    {\subfloat[\fontsize{7.4pt}{9pt}\selectfont Reconstructed samples of EuroSat dataset under LTP-FLEO when $L= 2$.]
    {\includegraphics[width=0.4\textwidth]{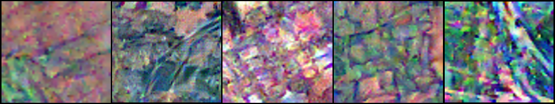}}}
    {\subfloat[\fontsize{7.4pt}{9pt}\selectfont Reconstructed samples of EuroSat dataset under LTP-FLEO when $L= 4$.]
    {\includegraphics[width=0.4\textwidth]{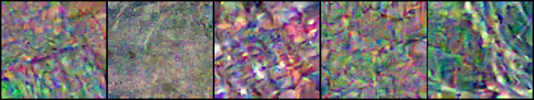}}}
    {\subfloat[\fontsize{7.4pt}{9pt}\selectfont Reconstructed samples of EuroSat dataset under LTP-FLEO when $L= 6$.]
    {\includegraphics[width=0.4\textwidth]{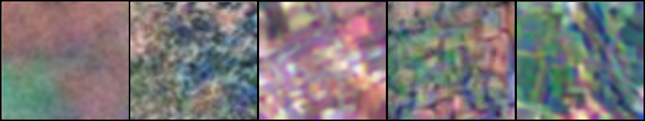}}}
{\caption{Comparing LTP-FLEO with FedSecure in reconstructing private satellite images under model inversion attack  by the $\mathcal {AS}$ on EuroSat dataset with various LTP Levels ($L$).\label{fig:EuroSat}}}\vspace{-0.1cm}
\end{figure}

{\bf Baselines.} To emphasize the significance of our proposed approach, LTP-FLEO, in protecting the privacy of satellites' models across multiple FL rounds and optimizing FL-LEO performance (in terms of convergence speed and accuracy), we conduct comparisons with the recent state-of-the-art. Firstly, we contrast LTP-FLEO with FedSecure~\cite{elmahallawy2023secure} in an asynchronous manner, a recent secure aggregation scheme for FL-LEO networks, where the $\mathcal{AS}$ updates the global model once it receives local models from any set of currently visible satellites to proceed to the next round. We furthermore assess LTP-FLEO against several FL-LEO approaches, including \cite{elmahallawy2022asyncfleo,elmahallawy2023secure,xie2020asynchronous}, to evaluate both the convergence speed and the achievable accuracy.

\subsection{Privacy-Preserving Results}

In \fref{fig:EuroSat}, we compare the reconstructed image resolutions achieved by our LTP-FLEO approach with those of FedSecure~\cite{elmahallawy2023secure}. Both methods utilize the same security scheme to encrypt satellite models. However, when the $\mathcal{AS}$ employs a model inversion attack using the historical of the global models—specifically the GradInversion technique~\cite{yin2021see}—FedSecure fails to preserve the privacy of the satellite images. In other words, the $\mathcal{AS}$ was able to successfully reconstruct private satellite images with high resolution, as shown in \ref{fig:EuroSat}.b. On the other hand, our LTP-FLEO approach demonstrates its robustness in preserving the satellite images' privacy over the long term. For instance, when the partition size is set to $L=2$, the $\mathcal{AS}$ failed to reconstruct individual satellite images (from which their labels could not be correctly identified), thus protecting the data privacy. As we increase $L$ to 4 and 6 (see \fref{fig:EuroSat}.d and e), the reconstructed images exhibit significantly lower resolutions, indicating a stronger LTP guarantee. However, this improvement comes at the cost of relatively slower FL convergence as we will delve into further in \sref{sec:Convergence}.
\begin{table}[!t]
\setlength{\tabcolsep}{0.6em}
\renewcommand{\arraystretch}{1.1}
\caption{Comparison of Image Reconstruction Performance: LTP-FLEO vs. FedSecure under various metrics.\vspace{-0.2cm}} 
\centering
\label{tab:2}
\begin{tabular}{p{2.05cm}|p{1.25cm}|>{\columncolor{lightlime}}p{1.25cm}|>{\columncolor{darklime}}p{1.25cm}|>{\columncolor{darkerlime}}p{1.25cm}} 
\bottomrule
Metric &FedSecure& \multicolumn{3}{c}{\bf LTP-FLEO}\\
\cline{2-5}
 &$L=1$ \newline(1.67 hour)&$L=2$\newline(1.86 hour) & $L=4$\newline(2.93 hour)  & $L=6$ \newline(4.56 hour)  \\
\hline 
PSNR $\downarrow$ & 10.68 & 5.28&3.03 & -1.98\\
\hline 
MSE $\uparrow$ & 0.1076&0.528 & 0.898& 1.577\\
\hline 
FMSE $\uparrow$ &0.1558 & 1.063& 1.926& 3.308\\
\hline
\footnotesize{Resolution Acc \%} &97.65 & 56.97& 36.16&22.96 \\
\toprule
\end{tabular}\vspace{-0.2cm}
\end{table}

In \tref{tab:2}, we also evaluate the reconstruction performance using four metrics: peak signal-to-noise ratio (PSNR), mean square error (MSE), feature-based MSE (FMSE), and resolution accuracy. A lower PSNR value indicates lower image quality with higher distortion compared to the original, implying a higher level of privacy preservation. Conversely, higher MSE and FMSE values indicate high similarity between the reconstructed and original images in terms of both pixel-wise and high-level features, respectively. The results presented in this table demonstrate that LTP-FLEO achieves lower PSNR and higher MSE and FMSE values at different partition sizes ($L=2,4,~\text{and}~6$) compared to FedSecure. This underscores the robustness of LTP-FLEO in ensuring LTP and shows that traditional secure aggregation schemes, such as FedSecure, reveal significant information about private satellite images.

\vspace{-0.25cm}
\subsection{Convergence and Fairness Results}\label{sec:Convergence}
\begin{figure}[!t]

    {\subfloat[ {MNIST -- IID setting.}]{\centering\includegraphics[width=0.24\textwidth]{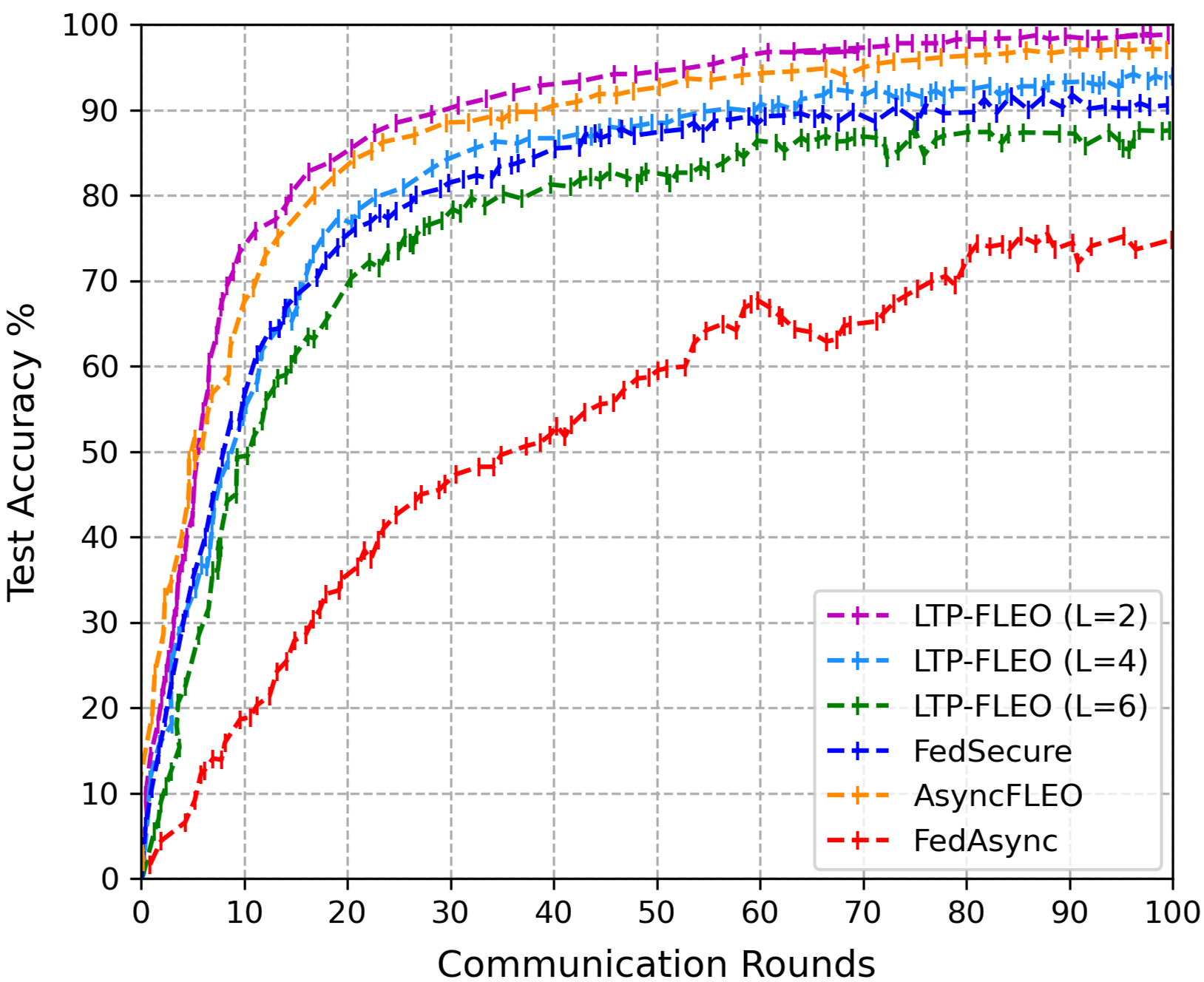}}}
    {\subfloat[{MNIST -- Non-IID setting.}]{\centering\includegraphics[width=0.24\textwidth]{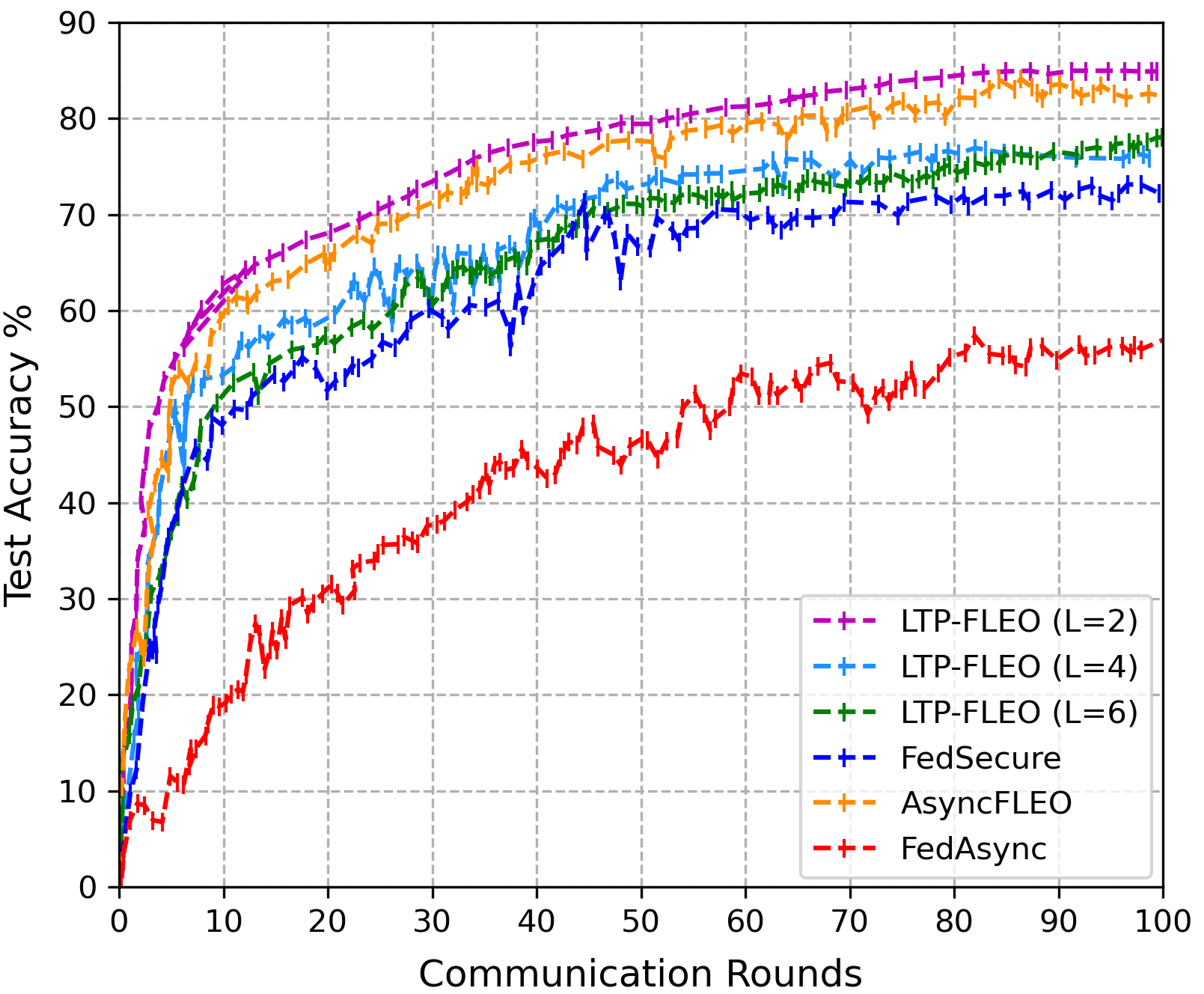}}}
    {\subfloat[{CIFAR-10 -- IID setting.}]{\centering\includegraphics[width=0.24\textwidth]{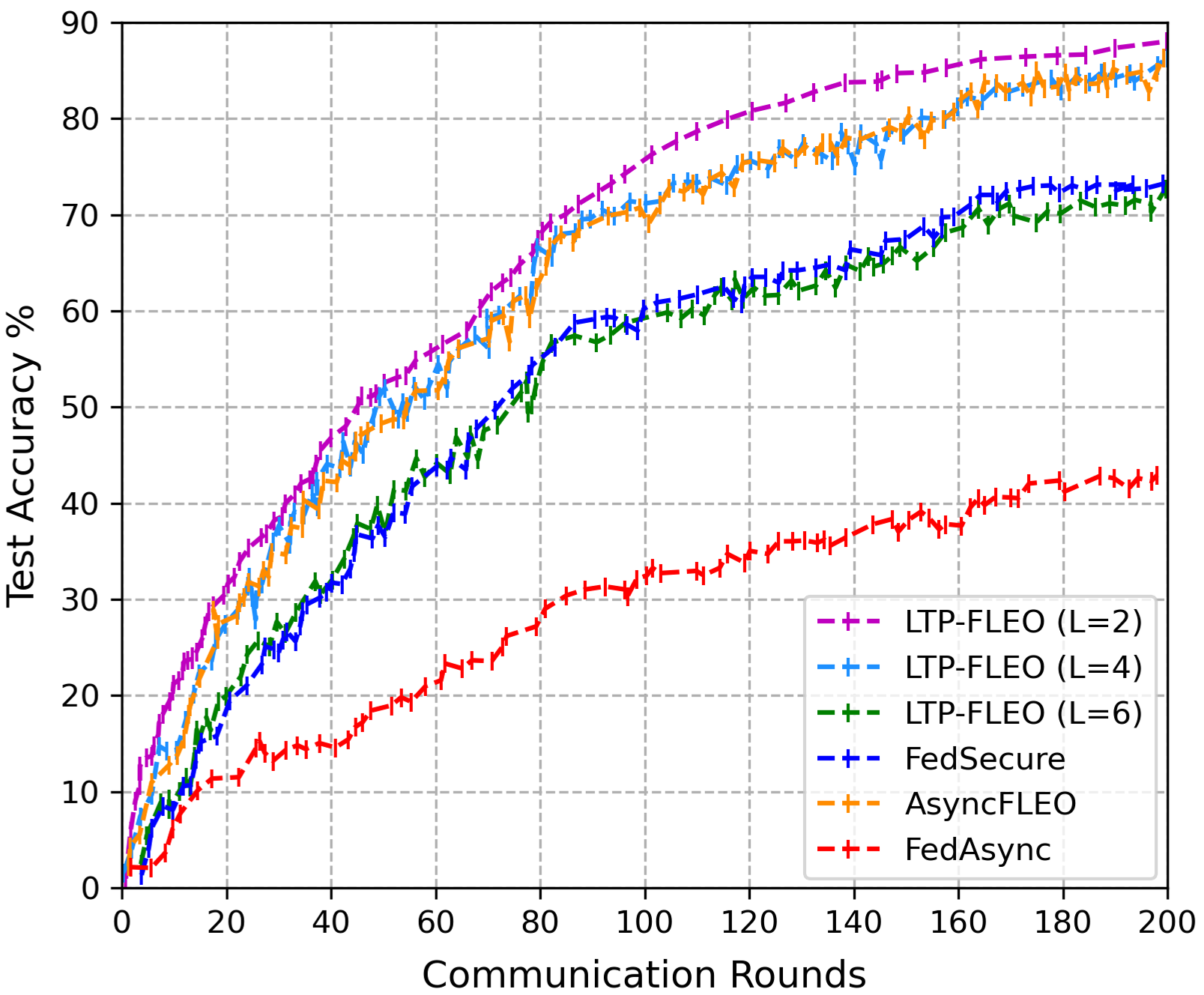}}}
    {\subfloat[ {CIFAR-10 -- non-IID setting.}]{\centering\includegraphics[width=0.24\textwidth]{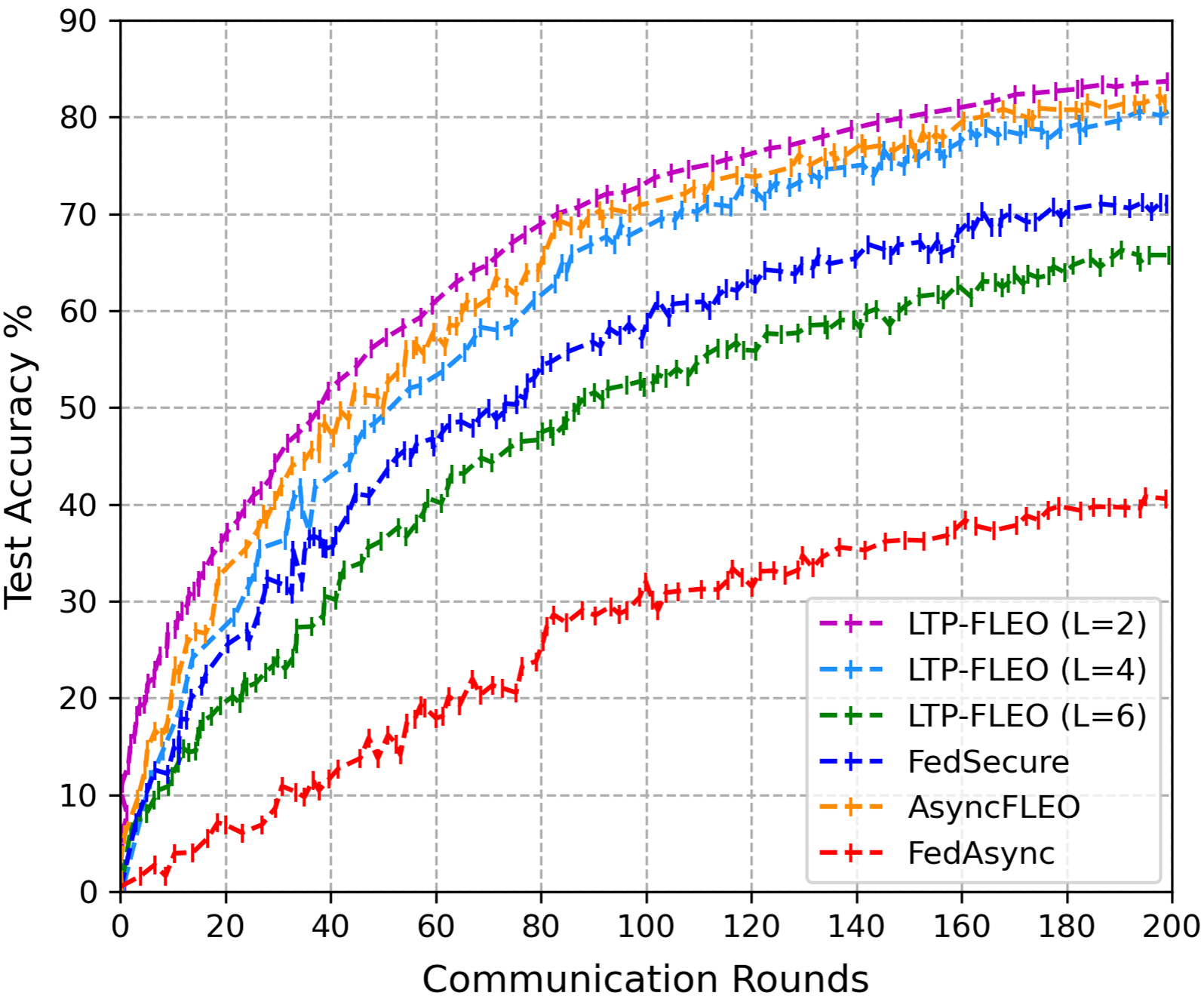}}}
{\caption{Performance comparison with baselines on MNIST and CIFAR-10 datasets under IID and non-IID data distributions.\label{fig:eval}}}
\end{figure}
{\bf Convergence Speed \& Accuracy Results.} We assess LTP-FLEO's convergence speed against baselines in two distinct data distribution settings: IID and non-IID. In non-IID settings, satellites in each orbit train ML models on only two image classes. We determine that each communication round for model collection and {\em global model update takes approximately 1-3 minutes, resulting in LTP-FLEO's convergence time spanning 1.86-4.56 hours} based on the target LTP guarantee (see \tref{tab:2}). 

In terms of convergence performance, LTP-FLEO achieves superior or at least comparable accuracy when benchmarked against existing baseline methods \cite{elmahallawy2022asyncfleo,elmahallawy2023secure,xie2020asynchronous} using datasets such as MNIST~\cite{deng2012mnist} and CIFAR-10~\cite{CIFAR-10}. As shown in Figure~\ref{fig:eval}, LTP-FLEO’s convergence accuracy is evaluated across different partition sizes ($L=2,4,6$) with a tolerance factor $\alpha=3$. The results demonstrate that LTP-FLEO (with $L=2$ and $L=4$) and AsyncFLEO~\cite{elmahallawy2022asyncfleo} achieve the highest convergence accuracy on the MNIST dataset under both IID and non-IID settings, ranging from 83.37\% to 96.62\%, with convergence times between 1.5-3.5 hours. When using the CIFAR-10 dataset, the accuracy slightly decreases to a range of 80.29\% to 84.13\%, and the convergence time roughly doubles to around 6.5 hours. In contrast, FedAsync~\cite{xie2020asynchronous} demonstrates the lowest convergence accuracy, ranging from 40.11\% to 73.65\% on both datasets. This performance degradation is primarily due to FedAsync’s inability to effectively handle the sporadic and irregular connectivity characteristic of LEO satellites with the $\mathcal{AS}$.


On the other hand, for LTP-FLEO, increasing the partition size to 6 slightly reduces convergence accuracy compared to partition sizes 2 and 4. This is because the waiting time for a larger number of satellites to transmit their models to the $\mathcal{AS}$ is longer than waiting for 2 or 4 satellites, which leads to extended communication times and slower convergence. As a result, this entails a trade-off between achieving optimal convergence performance and ensuring a more robust LTP guarantee.
\begin{figure}[!t]
\centering
    {\subfloat[\scriptsize {Prediction vs. Ground Truth.}]{\includegraphics[width= 0.21\textwidth]{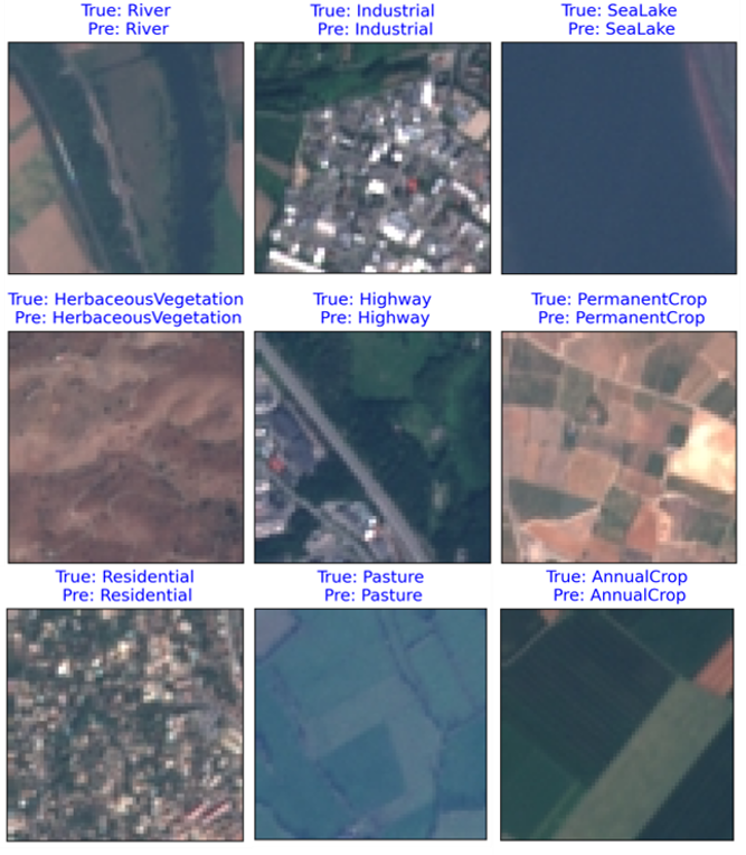}}}
    {\subfloat[\scriptsize {Confusion matrix across classes.}]{\includegraphics[width=0.26\textwidth]{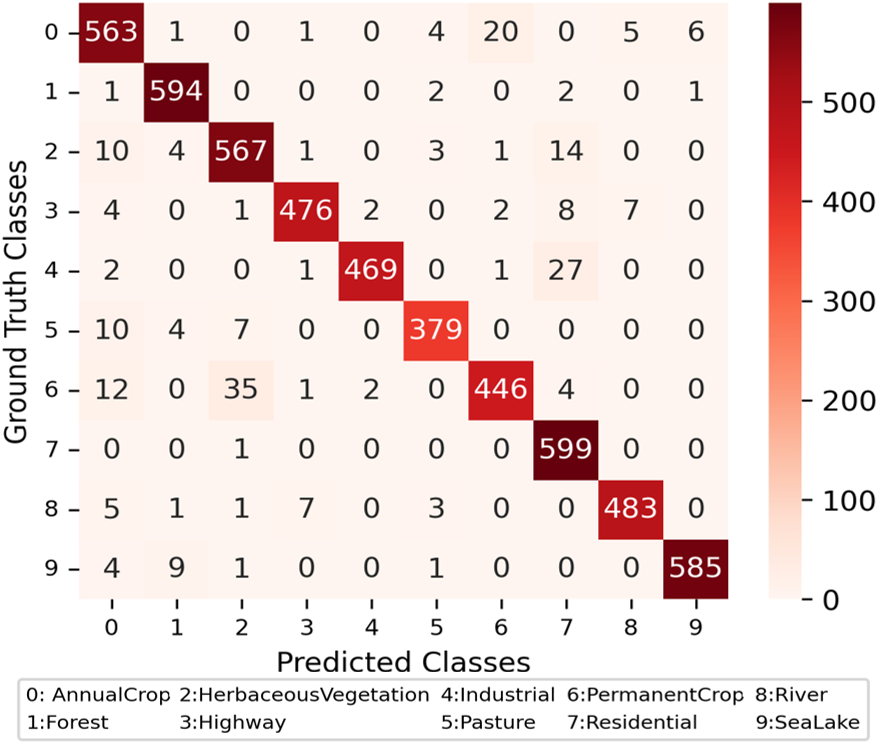}}}
{\caption{LTP-FLEO performance evaluation over EuroSat dataset.\label{fig:sample}}}
\end{figure}

\noindent{\bf Fairness Results.} To demonstrate the effectiveness of our fair global aggregation component, we investigate the performance of LTP-FLEO on the EuroSat dataset in a non-IID setting with $\alpha=t$. This means assessing the accuracy of the final global model in predicting diverse classes belonging to distinct satellites across various orbits. As shown in \fref{fig:sample}.a, LTP-FLEO accurately predicted all samples in this small batch, achieving an overall accuracy of approximately 95\% across all orbits and classes after only 5 hours. This achievement underscores the efficacy of the weighting factor $\beta$, which accounts for satellites' model staleness and data size, mitigating the adverse impact of their staleness on the global model. Moreover, depicted in \fref{fig:sample}.b, the confusion matrix highlights the correct predictions for each class. It reveals that the global model is unbiased towards specific classes over others, further emphasizing fair aggregation across satellites trained on diverse classes in distinct orbits.

\vspace{-3mm}
\section{Conclusion}\label{Sec:conculsion}
This paper presents LTP-FLEO, an asynchronous FL framework designed to preserve long-term privacy (LTP) in LEO satellite networks. LTP-FLEO safeguards the privacy of satellite models and their underlying data against both internal (curious) and external (malicious) threats. Unlike conventional secure aggregation schemes that provide only short-term privacy (STP) on a per-round basis, LTP-FLEO ensures LTP throughout the entire FL process.

The framework integrates three key components: privacy-aware satellite partitioning, model age balancing, and fair global aggregation. Our theoretical analysis and experimental evaluation---using both synthetic and real-world satellite datasets---confirm that LTP-FLEO effectively enforces LTP, achieves fast convergence, and maintains both high accuracy and fairness.

A promising future direction is to extend LTP-FLEO to cross-device {\em synchronous} FL settings. This could be realized by adapting the model age balancing mechanism to accommodate synchronized update schedules.

\section*{Appendix}
\appendix
\section{Proof of Theorem 3} 
The proof of Theorem 3 is structured in two parts. In the first part,  we provide the local update sequences for each satellite corresponding to each step index, and the updates for the global model $\boldsymbol{w}^t$ (as in Eq. (6)) for each global round index $t$. Subsequently, we provide the convergence expression for $\boldsymbol{w}^t$. In the second part, we provide the Big $\mathcal {O}$ notation for the necessary number of rounds $T$ for the global model $\boldsymbol{w}^t$ to attain the desired convergence accuracy.

\subsection{Convergence analysis of satellites' Local Models and the global model}

Let a satellite $k$ updates its local model $\boldsymbol{w}_{k}^{i}$ during local step $i$ with $I$ SGD local iterations, where $I\geq1$, before transmitting the updated local model version ($\boldsymbol{w}_{k}^{I}$ ) to the $\mathcal{AS}$. Additionally, let $\mathcal{I}_{I}$ represents the set of global synchronization steps, denoted as $\mathcal{I}_{I}= \{nI|n = 1, 2, \dots\}$ with the step-index $i$ increases from $nI$ to $nI+1$ only when the $\mathcal{AS}$ selects at least $\mathsf {K}$ satellite at step $nI + 1$ for global aggregation, where these satellites belong to any of the partitions in set $\mathcal G$. LTP-FLEO selects only $\mathcal U_{nI}$ satellites within partition/partitions that satisfy the LTP at step-index $nI$. Thus, the update equation of LTP-FLEO with dynamic in-contact satellites' visibility with the $\mathcal{AS}$, can be expressed as
\begin{align}\label{eq:new_var}
\boldsymbol{v}_{k}^{i+1}=\boldsymbol{w}_{k}^{i}- \eta_{i} \nabla F_{k} (\boldsymbol{w}_{k}^{i},\xi_{k}^{i}),\quad\eta_{k}^{i}={\{x_{j},y_{j}\}_{k}^{i}}
\end{align}
\begin{equation}
\boldsymbol{w}_{k}^{i+1}=
\begin{cases}
        \boldsymbol{v}_{k}^{i+1} \hspace{2cm}\text{if}~i+1 \notin \mathcal{I}_{I},\\ \frac{1}{\mathsf K}\sum_{k=1}^{\mathsf K}  \boldsymbol{v}_{k}^{i+1} \hspace{0.75cm}\text{if}~i+1 \in \mathcal{I}_{I}.
\end{cases}
\end{equation}

Here, in Equation~\eqref{eq:new_var}, an additional variable $\boldsymbol{v}_{k}^{i+1}$ is introduced, representing the immediate result of one step of SGD from $\boldsymbol{w}_{k}^{i}$. Additionally, we interpret $\boldsymbol{w}_{k}^{i+1}$ as the model parameters obtained after aggregation steps $i+1$ (if possible).

We introduce two virtual sequences 
motivated by \cite{ribero2022federated,li2019convergence}, namely,
\begin{align}
\boldsymbol{\Bar {v}}^{i}=\sum_{k=1}^{K}s_{k}\boldsymbol{\Bar v}_{k}^{i},\quad\boldsymbol{\Bar w}^{i}=\sum_{k=1}^{K}s_{k}\boldsymbol{\Bar w}_{k}^{i}.
\end{align}
Therefore, we can view $\boldsymbol{\Bar {v}}^{i+1}$ as  the outcome of a single SGD update step from $\boldsymbol{\Bar w}^{i}$. When $i+1$ falls outside $\mathcal{I}_I$, both $\boldsymbol{\Bar {v}}^{i}$ and $\boldsymbol{\Bar {w}}^{i}$ are not accessible. For convenience, we also define 
\begin{align}\label{eq:4}
\mathbf{\Bar g}^{i}&=\sum_{k=1}^{K}s_{k}\nabla F_{k}(\boldsymbol{w}_{k}^{i}),\\
\mathbf{g}^{i}&=\sum_{k=1}^{K}s_{k}\nabla F_{k}(\boldsymbol{w}_{k}^{i},\xi_{k}^{i}).  \nonumber  
\end{align}
According to the above Equation, Equation~\eqref{eq:4}, we have $\boldsymbol{\Bar {v}}^{i+1}=\boldsymbol{\Bar {w}}^{i}-\eta_{{i}} \mathbf{\Bar g}^{i}$

To ensure clarity in our proof, we establish several essential lemmas as outlined below.

\begin{lemma}[Fair Model Selection]~\\
If $i+1$  is within $\mathcal{I}_I$, $\boldsymbol{\Bar w}^{i+1}$ is accessible, denoted as  $$\mathbb{E}_{\mathcal U_{i+1}}[\boldsymbol{\Bar w}^{i+1}]=\boldsymbol{\Bar v}^{i+1}$$
\begin{proof}
Suppose $\mathcal U_{i+1}=\{k_1,\dots,k_{\mathsf K}\}$, and the sampling probability of each satellite is identical,,$Pr\{k=i\}=\frac{1}{K}$ for each $k\in 
\mathcal{K}$, reflecting the symmetry in the construction. Then, we can express the following relationship
\begin{align}
    \mathbb{E}_{\mathcal U_{i+1}}[\boldsymbol{\Bar w}^{i+1}]&=\frac{1}{\mathsf K}\mathbb{E}_{\mathcal U_{i+1}}\Bigg[\sum_{k=1}^{\mathsf K} \boldsymbol{v}_k^{i+1}\Bigg]\\
    &=\mathbb{E}_{\mathcal U_{i+1}}[\boldsymbol{ v}_k^{i+1}]=\sum_{k=1}^{K} \frac{1}{ K}\boldsymbol{v}_k^{i+1}=\boldsymbol{\Bar v}^{i+1} \nonumber
\end{align}
\end{proof}
\end{lemma}

\begin{lemma}[Convergence of one-step SGD] 
Suppose that Assumptions 1 and 2 hold. When $\eta_{{i}}\leq\frac{1}{4\mathcal L}$, then we have:
\begin{align}
     &\mathbb{E}\|\boldsymbol{\Bar {w}}^{i+1}-\boldsymbol{w}^*||^2_2\leq (1-\eta_{{i}}\mu)\mathbb{E}\|\boldsymbol{\Bar {w}}^{i}-\boldsymbol{w}^*||^2_2+\eta_{{i}}^2\mathbb{E}\|\mathbf{g}^{i}-\mathbf{\Bar g}^{i}\|^2_2\nonumber
     \\&+6\mathcal {L} \eta_{{i}}^2\Gamma+2\mathbb{E}\sum_{k=1}^{K}s_{k}\|\boldsymbol{\Bar {w}}^{i}-\boldsymbol{{w}}^{i}_{k}\|^2_2.   \nonumber
\end{align}
where $\boldsymbol{w}^*$ refers to the desired target model that can achieve the desired level of accuracy. 
\begin{proof}
   \begin{align}
      \|\boldsymbol{\Bar {w}}^{i+1}-\boldsymbol{w}^*||^2_2&=\|\boldsymbol{\Bar {w}}^{i+1}-\boldsymbol{\Bar {v}}^{i+1}+\boldsymbol{\Bar {v}}^{i+1}-\boldsymbol{w}^*||^2_2\\
       &=\underbrace{\|\boldsymbol{\Bar {w}}^{i+1}-\boldsymbol{\Bar {v}}^{i+1}||^2_2}_{A}+\underbrace{||\boldsymbol{\Bar {v}}^{i+1}-\boldsymbol{w}^*||^2_2}_{B}\nonumber\\
       &+\underbrace{2(\boldsymbol{\Bar {w}}^{i+1}-\boldsymbol{\Bar {v}}^{i+1})^\intercal (\boldsymbol{\Bar {v}}^{i+1}-\boldsymbol{w}^*)}_{C}\nonumber
   \end{align} 
According to Lemma 1, $\mathbb{E}_{\mathcal{U}_{i+1}}C$ will equal  zero. For the term $B$, we have
\begin{align}
    ||\boldsymbol{\Bar {v}}^{i+1}-\boldsymbol{w}^*||^2_2\leq(1-\eta_{{i}}\mu)\mathbb{E}\|\boldsymbol{\Bar {w}}^{i}-\boldsymbol{w}^*||^2_2+\eta_i^2\lambda
\end{align}

Suppose Assumption 3 holds, and the variance of $\mathbb{E}\|\mathbf{g}^{t}-\mathbf{\Bar g}^{t}\|^2_2$ is bounded,  such as $\mathbb{E}\|\mathbf{g}^{t}-\mathbf{\Bar g}^{t}\|^2_2\leq \sum_{k=1}^{K}s_{k}^2\sigma_{k}^{2}$. Consequently, the parameter $\lambda$ can be expressed as $\lambda=\sum_{k\in \mathcal{K}} {{s_k}}^2{\sigma_{k}}^2+6\mathcal{L} \Gamma+8(I-1)^2H^2$.

Suppose Assumption 4 also holds, and $\eta_{i}\leq2\eta_{i+I}$ is non-increasing $\forall~i\geq 1$, then the divergence of term $A$ is bounded as
\begin{equation} \mathbb{E}_{i+1}\Bigg[\sum_{k=1}^{K}s_{k}\|\boldsymbol{{\Bar w}}^{i}_{k}-\boldsymbol{{w}}^{i}_{k}\|^2_2\Bigg]\leq\eta_{{i}}^2\nu
\end{equation}
where $\nu=\frac{4I^2 H^2}{K}$.
\end{proof}
\end{lemma}
From Lemma 1 and Lemma 2, we have
\begin{align}
    \mathbb{E}\|\boldsymbol{\Bar {w}}^{i+1}-\boldsymbol{w}^*||^2_2&\Resize{5.3cm}{\leq (1-\eta_{{i}}\mu)\mathbb{E}\|\boldsymbol{\Bar {w}}^{i}-\boldsymbol{w}^*||^2_2+\eta_i^2(\lambda+\nu)}
\end{align}
By employing the induction as in~\cite{li2019convergence}, we can demonstrate that
\begin{align}
     \mathbb{E}\|\boldsymbol{\Bar {w}}^{i+1}-\boldsymbol{w}^*||^2_2
    &\Resize{5.2cm}{\leq\frac{1/\mu}{\upsilon+i}\Bigg(\frac{2(\lambda+\nu)}{\mu}+\frac{\mu\upsilon}{2}\mathbb{E}\|\boldsymbol{\Bar {w}}^{1}-\boldsymbol{w}^*||^2_2\Bigg)}
\end{align}
where $\upsilon=\max{8\kappa,I}$ and $\kappa=\frac{\mathcal L}{\mu}$. Leveraging Assumption 1, which asserts the strong $\mathcal{L}$-smoothness of $F(\cdot)$, then Equation~(10) can be rewritten as
\begin{align} 
\mathbb{E} [F(\boldsymbol{\Bar{w}}^{I})]-F^*& \Resize{5.5cm}{\leq\frac{\mathcal{L}/\mu}{\upsilon+I}\Bigg(\frac{2(\lambda+\nu)}{\mu}+\frac{\mu\upsilon}{2}\mathbb{E}\|\boldsymbol{\Bar {w}}^{1}-\boldsymbol{w}^*||^2_2\Bigg)}\\
    &\Resize{5.5cm}{\leq\frac{2\kappa}{\upsilon+I}\Bigg(\frac{(\lambda+\nu)}{\mu}+\frac{\mu\upsilon}{4}\mathbb{E}\|\boldsymbol{\Bar {w}}^{1}-\boldsymbol{w}^*||^2_2\Bigg)\nonumber}
\end{align}

To provide the convergence analysis of the global model$\boldsymbol{\Bar{w}}^{T} $ over $T$ global rounds, since we define the step-index $i$ increases from $nI$ to $nI + 1$ only when the $\mathcal {AS}$ can select $\mathsf {K}$, we have:
\begin{align} 
\mathbb{E} [F(\boldsymbol{{w}}^{T})]=\mathbb{E} [F(\boldsymbol{\Bar{w}}^{(IT\phi)})]
\end{align}
Here $\phi$ denotes the probability of having at least $\mathsf {K}$ satellites in-contact with $\mathcal {AS}$, available for global aggregation, at each synchronization step. By comparing Equation (11) and (12), we obtain:
\begin{align} 
\Resize{7.6cm}{\mathbb{E} [F(\boldsymbol{{w}}^{T})]-F^*\leq\frac{2\kappa}{\upsilon+IT}\Bigg(\frac{(\lambda+\nu)}{\mu}+\frac{\mu\upsilon}{4}\mathbb{E}\|\boldsymbol{{w}}^{1}-\boldsymbol{w}^*||^2_2\Bigg)}
\end{align}

\subsection{The upper bound of global rounds for global model convergence}

Our LTP-FLEO approach requires two iterations (as traditional Federated Learning approaches)—one for broadcasting the global model and another for aggregation—during each complete communication round. When the total number of iterations is sufficiently large, we can establish an upper bound on the required number of communication rounds as
\begin{align}
\Resize{7.6cm}{T/I\propto \Big(1+\frac{1}{K}\Big)IH^2+\frac{\sum_{k=1}^{ K}s_k^2\sigma_k^2+\mathcal{L}\Gamma+\kappa H^2}{I}+H^2,}
\end{align}
This indicates that the optimal local step $I^*$ exists, as the function of $I$ that first decreases and then increases. Furthermore, the value of $T/I$  can be evaluated at $I^*$ as
\begin{align}
    \Resize{5cm}{\mathcal{O}\Bigl(H\sqrt{\sum_{k=1}^{ K}s_k^2\sigma_k^2+\mathcal{L}\Gamma+\kappa H^2}~\Bigl),}
\end{align}
Equation (15) shows that LTP-FLEO requires more communication rounds to address more severe heterogeneity among satellites.

Without loss of generality, by utilizing Equations (14) and (15), we can express the required number of rounds for LTP-FLEO to achieve a certain accuracy $\rho$ of the global model $\boldsymbol{{w}}^{T}$ when satellites are partitioned into a set of partitions $\mathcal G$ as follows:
\begin{align}
\Resize{7.5cm}{{T=\mathcal{O}\Bigg(\frac{1}{\rho}\biggl(\Big(1+\frac{1}{|\mathcal{G}|}\Big)IH^2+\frac{\sum_{k\in\mathcal{G}}s_k^2\sigma_k^2+\mathcal{L}\Gamma+\kappa H^2}{I}}+H^2\biggl)\Bigg)}
\end{align}



\bibliography{mybibfile}

\begin{thebibliography}{37}
\providecommand{\natexlab}[1]{#1}
\providecommand{\url}[1]{\texttt{#1}}
\expandafter\ifx\csname urlstyle\endcsname\relax
  \providecommand{\doi}[1]{doi: #1}\else
  \providecommand{\doi}{doi: \begingroup \urlstyle{rm}\Url}\fi

\bibitem[Ali et~al.(1999)Ali, Al-Dhahir, and Hershey]{ali1999predicting}
I.~Ali, N.~Al-Dhahir, and J.~E. Hershey.
\newblock Predicting the visibility of leo satellites.
\newblock \emph{IEEE Transactions on Aerospace and Electronic Systems}, 35\penalty0 (4):\penalty0 1183--1190, 1999.

\bibitem[{Ansys Inc.}(2024)]{ansys_stk}
{Ansys Inc.}
\newblock {Ansys STK - Systems Tool Kit}, 2024.
\newblock URL \url{https://www.ansys.com/products/missions/ansys-stk}.
\newblock Accessed: 2025-04-14.

\bibitem[Bonawitz et~al.(2017)Bonawitz, Ivanov, Kreuter, Marcedone, McMahan, Patel, Ramage, Segal, and Seth]{bonawitz2017practical}
K.~Bonawitz, V.~Ivanov, B.~Kreuter, A.~Marcedone, H.~B. McMahan, S.~Patel, D.~Ramage, A.~Segal, and K.~Seth.
\newblock Practical secure aggregation for privacy-preserving machine learning.
\newblock In \emph{proceedings of the 2017 ACM SIGSAC Conference on Computer and Communications Security}, pages 1175--1191, 2017.

\bibitem[Chang et~al.(2023)Chang, Zhang, Gong, and Qian]{chang2023privacy}
Y.~Chang, K.~Zhang, J.~Gong, and H.~Qian.
\newblock Privacy-preserving federated learning via functional encryption, revisited.
\newblock \emph{IEEE Transactions on Information Forensics and Security}, 18:\penalty0 1855--1869, 2023.

\bibitem[Chen et~al.(2022{\natexlab{a}})]{chen2022fundamental}
W.-N. Chen et~al.
\newblock The fundamental price of secure aggregation in differentially private federated learning.
\newblock In \emph{International Conference on Machine Learning}, pages 3056--3089. PMLR, 2022{\natexlab{a}}.

\bibitem[Chen et~al.(2022{\natexlab{b}})Chen, Zhang, Li, Che, Jin, Guo, Yang, An, and Nie]{IoRT2022tibet}
Y.~Chen, M.~Zhang, X.~Li, T.~Che, R.~Jin, J.~Guo, W.~Yang, B.~An, and X.~Nie.
\newblock Satellite-enabled internet of remote things network transmits field data from the most remote areas of the tibetan plateau.
\newblock \emph{Sensors}, 22\penalty0 (10):\penalty0 3713, 2022{\natexlab{b}}.
\newblock \doi{10.3390/s22103713}.

\bibitem[Deng(2012)]{deng2012mnist}
L.~Deng.
\newblock The mnist database of handwritten digit images for machine learning research.
\newblock \emph{IEEE Signal Processing Magazine}, 29\penalty0 (6):\penalty0 141--142, 2012.

\bibitem[Dong et~al.(2023)Dong, Xiaojun, Jing, Kaiyun, and Wang]{dong2023meteor}
Y.~Dong, C.~Xiaojun, W.~Jing, L.~Kaiyun, and W.~Wang.
\newblock Meteor: improved secure 3-party neural network inference with reducing online communication costs.
\newblock In \emph{Proceedings of the ACM Web Conference 2023}, pages 2087--2098, 2023.

\bibitem[Elmahallawy and Luo(2022)]{elmahallawy2022asyncfleo}
M.~Elmahallawy and T.~Luo.
\newblock {AsyncFLEO}: Asynchronous federated learning for {LEO} satellite constellations with high-altitude platforms.
\newblock In \emph{2022 IEEE International Conference on Big Data (Big Data)}, pages 5478--5487. IEEE, 2022.

\bibitem[Elmahallawy et~al.(2023)Elmahallawy, Luo, and Ibrahem]{elmahallawy2023secure}
M.~Elmahallawy, T.~Luo, and M.~I. Ibrahem.
\newblock Secure and efficient federated learning in {LEO} constellations using decentralized key generation and on-orbit model aggregation.
\newblock In \emph{IEEE Global Communications Conference (GLOBECOM)}, 2023.

\bibitem[Elmahallawy et~al.(2024)Elmahallawy, Luo, and Ramadan]{10438925}
M.~Elmahallawy, T.~Luo, and K.~Ramadan.
\newblock Communication-efficient federated learning for leo satellite networks integrated with haps using hybrid noma-ofdm.
\newblock \emph{IEEE Journal on Selected Areas in Communications}, pages 1--1, 2024.

\bibitem[Fang et~al.(2023)Fang, Zhai, Yu, Wu, Gong, and Chen]{IoRT2023olive}
Q.~Fang, Z.~Zhai, S.~Yu, Q.~Wu, X.~Gong, and X.~Chen.
\newblock Olive branch learning: A topology-aware federated learning framework for space-air-ground integrated network.
\newblock \emph{IEEE Transactions on Wireless Communications}, 22\penalty0 (7):\penalty0 4534--51, July 2023.

\bibitem[Geiping et~al.(2020)]{geiping2020inverting}
J.~Geiping et~al.
\newblock Inverting gradients-how easy is it to break privacy in federated learning?
\newblock \emph{Advances in Neural Information Processing Systems (NeurIPS)}, 33:\penalty0 16937--16947, 2020.

\bibitem[Hassan et~al.(2023)]{hassan2023sfl}
S.~S. Hassan et~al.
\newblock Sfl-leo: Secure federated learning computation based on leo satellites for 6g non-terrestrial networks.
\newblock In \emph{NOMS 2023-2023 IEEE/IFIP Network Operations and Management Symposium}, pages 1--5. IEEE, 2023.

\bibitem[Helber et~al.(2019)Helber, Bischke, Dengel, and Borth]{helber2019eurosat}
P.~Helber, B.~Bischke, A.~Dengel, and D.~Borth.
\newblock Eurosat: A novel dataset and deep learning benchmark for land use and land cover classification.
\newblock \emph{IEEE Journal of Selected Topics in Applied Earth Observations and Remote Sensing}, 12\penalty0 (7):\penalty0 2217--2226, 2019.

\bibitem[Kanagavelu et~al.(2020)]{kanagavelu2020two}
R.~Kanagavelu et~al.
\newblock Two-phase multi-party computation enabled privacy-preserving federated learning.
\newblock In \emph{2020 20th IEEE/ACM International Symposium on Cluster, Cloud and Internet Computing (CCGRID)}, pages 410--419. IEEE, 2020.

\bibitem[Krizhevsky et~al.(2010)Krizhevsky, Nair, and Hinton]{CIFAR-10}
A.~Krizhevsky, V.~Nair, and G.~Hinton.
\newblock Cifar-10 (canadian institute for advanced research).
\newblock \emph{URL http://www. cs. toronto. edu/kriz/cifar. html}, 5\penalty0 (4):\penalty0 1, 2010.

\bibitem[Li et~al.(2021)Li, He, and Song]{lipractical}
Q.~Li, B.~He, and D.~Song.
\newblock Practical one-shot federated learning for cross-silo setting.
\newblock In \emph{Int. Joint Conf. on Artificial Intelligence (IJCAI)}, 2021.

\bibitem[Li et~al.(2019)Li, Huang, Yang, Wang, and Zhang]{li2019convergence}
X.~Li, K.~Huang, W.~Yang, S.~Wang, and Z.~Zhang.
\newblock On the convergence of fedavg on non-iid data.
\newblock \emph{arXiv preprint arXiv:1907.02189}, 2019.

\bibitem[Li et~al.(2020)Li, Huang, Yang, Wang, and Zhang]{Li2020On}
X.~Li, K.~Huang, W.~Yang, S.~Wang, and Z.~Zhang.
\newblock On the convergence of fedavg on non-iid data.
\newblock In \emph{International Conference on Learning Representations}, 2020.
\newblock URL \url{https://openreview.net/forum?id=HJxNAnVtDS}.

\bibitem[Lin et~al.(2023)Lin, Chen, Fang, Chen, Wang, and Gao]{lin2023fedsn}
Z.~Lin, Z.~Chen, Z.~Fang, X.~Chen, X.~Wang, and Y.~Gao.
\newblock Fedsn: A general federated learning framework over {LEO} satellite networks.
\newblock \emph{arXiv preprint arXiv:2311.01483}, 2023.

\bibitem[Liu et~al.(2023)Liu, Lin, and Liu]{liu2023long}
Z.~Liu, H.-Y. Lin, and Y.~Liu.
\newblock Long-term privacy-preserving aggregation with user-dynamics for federated learning.
\newblock \emph{IEEE Transactions on Information Forensics and Security}, 2023.

\bibitem[Ma et~al.(2022)Ma, Naas, Sigg, and Lyu]{ma2022privacy}
J.~Ma, S.-A. Naas, S.~Sigg, and X.~Lyu.
\newblock Privacy-preserving federated learning based on multi-key homomorphic encryption.
\newblock \emph{International Journal of Intelligent Systems}, 37\penalty0 (9):\penalty0 5880--5901, 2022.

\bibitem[McMahan et~al.(2017)McMahan, Moore, Ramage, Hampson, and y~Arcas]{mcmahan2017communication}
B.~McMahan, E.~Moore, D.~Ramage, S.~Hampson, and B.~A. y~Arcas.
\newblock Communication-efficient learning of deep networks from decentralized data.
\newblock In \emph{AISTATS}, pages 1273--1282, 2017.

\bibitem[Nasr et~al.(2019)Nasr, Shokri, and Houmansadr]{nasr2019comprehensive}
M.~Nasr, R.~Shokri, and A.~Houmansadr.
\newblock Comprehensive privacy analysis of deep learning: Passive and active white-box inference attacks against centralized and federated learning.
\newblock In \emph{2019 IEEE symposium on security and privacy (SP)}, pages 739--753. IEEE, 2019.

\bibitem[{\"O}stman et~al.(2023){\"O}stman, Gomez, Shreenath, and Meoni]{ostman2023decentralised}
J.~{\"O}stman, P.~Gomez, V.~M. Shreenath, and G.~Meoni.
\newblock Decentralised semi-supervised onboard learning for scene classification in low-earth orbit.
\newblock \emph{arXiv preprint arXiv:2305.04059}, 2023.

\bibitem[Ribero et~al.(2022)Ribero, Vikalo, and De~Veciana]{ribero2022federated}
M.~Ribero, H.~Vikalo, and G.~De~Veciana.
\newblock Federated learning under intermittent client availability and time-varying communication constraints.
\newblock \emph{IEEE Journal of Selected Topics in Signal Processing}, 17\penalty0 (1):\penalty0 98--111, 2022.

\bibitem[Rossi et~al.(2025)Rossi, De~Souza, and Luizelli]{rossi2025resource}
F.~D. Rossi, P.~S.~S. De~Souza, and M.~C. Luizelli.
\newblock Resource allocation on low-earth orbit edge infrastructure: Taxonomy, survey, and research challenges.
\newblock \emph{IEEE Access}, 2025.

\bibitem[Routray et~al.(2020)Routray, Javali, Sahoo, Sharmila, and Anand]{routray2020military}
S.~K. Routray, A.~Javali, A.~Sahoo, K.~Sharmila, and S.~Anand.
\newblock Military applications of satellite based {IoT}.
\newblock In \emph{2020 Third International Conference on Smart Systems and Inventive Technology (ICSSIT)}, pages 122--127. IEEE, 2020.

\bibitem[Shi et~al.(2024)Shi, Zeng, Zhu, Zhou, Jiang, and Letaief]{shi2024satellite}
Y.~Shi, L.~Zeng, J.~Zhu, Y.~Zhou, C.~Jiang, and K.~B. Letaief.
\newblock Satellite federated edge learning: Architecture design and convergence analysis.
\newblock \emph{IEEE Transactions on Wireless Communications}, 2024.

\bibitem[So et~al.(2023)So, Ali, G{\"u}ler, Jiao, and Avestimehr]{so2023securing}
J.~So, R.~E. Ali, B.~G{\"u}ler, J.~Jiao, and A.~S. Avestimehr.
\newblock Securing secure aggregation: Mitigating multi-round privacy leakage in federated learning.
\newblock In \emph{Proceedings of the AAAI Conference on Artificial Intelligence}, volume~37, pages 9864--9873, 2023.

\bibitem[So et~al.(2021)]{sosecure}
J.~So et~al.
\newblock Secure aggregation for buffered asynchronous federated learning.
\newblock In \emph{NeurIPS Workshop on New Frontiers in Federated Learning (NFFL)}, 2021.

\bibitem[Wei et~al.(2020)Wei, Li, Ding, Ma, Yang, Farokhi, Jin, Quek, and Poor]{wei2020federated}
K.~Wei, J.~Li, M.~Ding, C.~Ma, H.~H. Yang, F.~Farokhi, S.~Jin, T.~Q. Quek, and H.~V. Poor.
\newblock Federated learning with differential privacy: Algorithms and performance analysis.
\newblock \emph{IEEE Transactions on Information Forensics and Security}, 15:\penalty0 3454--3469, 2020.

\bibitem[Wu et~al.(2023)Wu, Zhang, and Luo]{wu2023esafl}
J.~Wu, W.~Zhang, and F.~Luo.
\newblock Esafl: Efficient secure additively homomorphic encryption for cross-silo federated learning.
\newblock \emph{arXiv preprint arXiv:2305.08599}, 2023.

\bibitem[Xiang et~al.(2024)Xiang, Ioannidis, Yeh, Joe-Wong, and Su]{xiang2024efficient}
M.~Xiang, S.~Ioannidis, E.~Yeh, C.~Joe-Wong, and L.~Su.
\newblock Efficient federated learning against heterogeneous and non-stationary client unavailability.
\newblock \emph{Advances in Neural Information Processing Systems}, 37:\penalty0 104281--104328, 2024.

\bibitem[Xie et~al.(2020)]{xie2020asynchronous}
C.~Xie et~al.
\newblock Asynchronous federated optimization.
\newblock In \emph{12$^{th}$ Wksp on Optimization for Machine Learning}, 2020.

\bibitem[Yin et~al.(2021)Yin, Mallya, Vahdat, Alvarez, Kautz, and Molchanov]{yin2021see}
H.~Yin, A.~Mallya, A.~Vahdat, J.~M. Alvarez, J.~Kautz, and P.~Molchanov.
\newblock See through gradients: Image batch recovery via gradinversion.
\newblock In \emph{Proceedings of the IEEE/CVF Conference on Computer Vision and Pattern Recognition}, pages 16337--16346, 2021.

\end{thebibliography}

\end{document}